\documentclass[article,12pt]{article}
\pdfoutput=1

\usepackage{jheppub} 

\usepackage{amssymb,amsbsy,amsfonts,amsmath}
\usepackage{epsfig}
\usepackage{cancel}
\usepackage{color}

\usepackage{subfigure}
\usepackage{graphicx}
\usepackage{epstopdf}
\usepackage{float}

\def\bea{\begin{eqnarray}}
\def\eea{\end{eqnarray}}
\def\be{\begin{equation}}
\def\ee{\end{equation}}
\def\beqn{\begin{eqnarray}}
\def\eeqn{\end{eqnarray}}
\def\beq{\begin{equation}}
\def\eeq{\end{equation}}

\def\Dslash{\not{\hbox{\kern-4pt $D$}}}
\def\pslash{\not{\hbox{\kern-4pt $p$}}}

\begin{document}

\baselineskip=21pt \pagestyle{plain} \setcounter{page}{1}

\vspace*{-1.7cm}

\begin{flushright}{\small Fermilab-PUB-14-386-T}\end{flushright}

\vspace*{0.2cm}
 
 \begin{center}

{\large \bf   GeV-scale dark matter: production at the Main Injector} \\ [9mm]

{\normalsize \bf Bogdan~A.~Dobrescu  and Claudia~Frugiuele } \\ [4mm]
 {\it Theoretical Physics Department, Fermilab, Batavia, IL 60510, USA } 


%

\vspace*{0.5cm}

October 6, 2014

\vspace*{0.7cm}

{\bf \small Abstract}

\vspace*{0.4cm}

\parbox{14.8cm}{Assuming that dark matter particles interact with quarks via a GeV-scale mediator, we study dark matter production in fixed target collisions.
The ensuing signal in a neutrino near detector consists of neutral-current events with an energy distribution peaked at higher values than the neutrino background.
We find that for a $Z'$ boson of mass around a few  GeV that decays to dark matter particles,  
the dark matter beam produced by the Main Injector at Fermilab allows the exploration of a range of values for the gauge coupling 
that currently satisfy all experimental constraints. The NO$\nu$A detector is well positioned for probing the presence of a dark matter beam, while future 
LBNF near-detectors would provide more sensitive probes.
}

\end {center}

\tableofcontents

\section{Introduction}

Dark matter (DM) provides solid evidence for physics beyond the Standard Model (SM), but its identity remains unknown.
A central question is whether DM particles experience interactions with ordinary matter beyond gravity.
If DM particles have weak-scale masses and order-one couplings, then their relic abundance given by the thermal freeze-out is consistent with the observed value \cite{Agashe:2014kda}.
An extensive experimental program has been carried out to explore this WIMP  (weakly-interacting massive particles) paradigm, setting impressive bounds on the viable parameter space.
Direct detection experiments \cite{Cushman:2013zza}, in particular, have imposed strong constraints on the interactions between nucleons and WIMPs of mass larger than about 5 GeV. 

Here we consider the possibility that DM particles have mass below 5 GeV and have interactions with quarks mediated by a new boson. 
If the latter is relatively light, then the DM interactions with nucleons can be probed in experiments.
A natural choice for the mediator is a leptophobic $Z'$ boson of mass near the GeV scale. The constraints on a boson of this type interacting with quarks are rather loose \cite{Dobrescu:2014fca,Tulin:2014tya}.
We will show that the constraints remain weak even when the $Z'$ interacts with DM.

A promising search method for such light DM particles is to test if they are 
produced in fixed target experiments, which benefit from large beam intensities.
We will focus on the case where the $Z'$ boson decays into a pair of DM particles.
Proton scattering off a fixed target may copiously  produce  $Z'$ bosons of mass around a few GeV 
as long as the proton energy is larger than a few tens of GeV. 
Thus, the Main Injector \cite{Anderson:1998zza} at Fermilab, which accelerates protons at 120 GeV, is well suited to test this GeV-scale DM scenario. In the NuMi beam line, where the 120 GeV protons hit a Carbon target, 
if DM particles  are produced, then they could be detected as neutral-current events in neutrino near-detectors such as NO$\nu$A \cite{Patterson:2012zs,O'ShegOshinowo:2013naa} and MINOS  \cite{MINOS:1998nfa,Thron:1996vc}. The future LBNF  \cite{Papadimitriou:2014raa} beam line would be even better suited 
for this type of search, provided a detector is placed within a few hundred meters from the target.

The possibility of searching for dark matter beams at neutrino detectors  has been recently proposed and explored in \cite{Batell:2009di, deNiverville:2011it,deNiverville:2012ij, Batell:2014yra}, especially for DM much lighter than 1 GeV, which can  
be produced in the Fermilab Booster beam line where the protons have an energy of 8 GeV \cite{Dharmapalan:2012xp}. 
Hadron collider experiments are also sensitive to light DM that interacts with quarks, because a pair of DM particles could be produced in association with a jet or a photon or other particles \cite{Goodman:2010yf,Bai:2010hh,Goodman:2010ku,Fox:2011pm,Goodman:2011jq, An:2012va, Fox:2012ee,An:2012ue}.
Quarkonium decays that involve missing energy provide another probe of the GeV-scale DM scenario \cite{Fayet:2006sp,Fayet:2007ua,Fayet:2009tv,Graesser:2011vj,Shoemaker:2011vi}.
We are going to compare the existing limits from these classes of experiments, and show that DM beams produced at the Main Injector may lead to thousands of deep-inelastic 
neutral current events in existing and future neutrino detectors.
This is an example of a broader capability of high-intensity fixed target experiments to probe the existence of light hidden particles   \cite{Reece:2009un,Bjorken:2009mm,Essig:2010xa,Essig:2010gu,Izaguirre:2013uxa,Morrissey:2014yma,Izaguirre:2014dua}.



%

\section{Leptophobic $Z'$ as portal to hidden particles}
\label{sec:}\setcounter{equation}{0}

We are focusing on a vector boson $Z'$ of mass $M_{Z'}$ in the $1-10$ GeV range, so that it can 
be produced by an $O(100)$ GeV proton beam scattering off a fixed target. 
To that end, we extend the SM gauge group by including an $U(1)_z$ group under which the quarks are charged while the leptons are neutral.
The simplest charge assignment that allows quark masses and 
evades constraints from flavor-changing neutral currents (FCNC)
is  charges given by the baryon number, $U(1)_B$ \cite{Nelson:1989fx,Carone:1994aa,Carone:1995pu, Dobrescu:2013cmh,Duerr:2013dza}.
Another possibility is to assign charge 0 to the left-handed quark doublets and to charge either the down- or the up-type right-handed quarks; a simple choice 
is $U(1)_{ds}$, where $d_R$ and $s_R$  have charges +1 and $-1$, while all other quarks are neutral \cite{Dobrescu:2014fca}.
Thus,  the $Z'$ boson has the following couplings to the SM quarks $q = u,d,s,c,b,t$,
\be
\mathcal{L}_q = \frac{g_z}{2}  Z^{\prime \mu} \times \left\{ 
\begin{array}{c}   \, {\displaystyle \frac{1}{3} \sum_q } \; \overline q \gamma_\mu q  \; ,  \;\; \;  U(1)_B  {\rm \;\; case }    ~~,
\\ [3mm] 
\, \overline d_R \gamma_\mu d_R - \overline s_R \gamma^\mu s_R 
\; , \;\; \;  U(1)_{ds}  {\rm \;\; case }   ~~,
\end{array} \right.
\label{MinimalModel}
\ee

Let us also include a very long lived particle, generically labelled by $\chi$, of mass $m_\chi < M_{Z'}/2$, which is a 
color singlet, electrically neutral, but charged under $U(1)_z$.  We focus on the cases where $\chi$  is either a Dirac fermion or a complex scalar (possible DM scenarios are discussed in Section 3).
Occasionally we will use the notations $ \psi_{\chi} $ or $\phi_{\chi} $ when we need to emphasize the difference between the fermion and scalar $\chi$.
If $\chi$ is a Dirac fermion, then 
its left- and right-handed components may have different $U(1)_z$  charges; for simplicity we will ignore this possibility though, and label 
the $U(1)_z$ charge of $\chi$ by $z_\chi$ whether it  is a fermion or a complex scalar. The $Z'$ couplings to the long-lived particle $\chi$ are
\be
\mathcal{L}_\chi = \frac{g_z}{2}  Z^{\prime \mu} \times \left\{ 
\begin{array}{c}  z_\chi \overline \psi_\chi \gamma_\mu \psi_\chi    \; ,  \;\; \;   {\rm if \;  Dirac \;  fermion \;\; }   ~~,
\\ [3mm] i z_\chi \left[ (\partial_\mu \phi_\chi^\dagger) \phi_\chi  -  \phi_\chi^\dagger  \partial_\mu \phi_\chi \right] 
 \; ,  \;\; \;   {\rm if \;  complex \;  scalar \;\; }   ~~,
 \end{array} \right.
\ee

The partial width for the $Z'$ decay into a pair of $\chi$ particles is
\bea
 \Gamma ( Z' \rightarrow  \psi_{\chi}  \bar   \psi_{\chi} ) =  \frac{g_z^2 z_{\chi}^2}{48 \pi} M_{Z'}  \Big(1+2 \frac{ m_{\chi}^2}{M_{Z'}^2 } \Big) 
  \Big(1-4 \frac{ m_{\chi}^2}{M_{Z'}^2 } \Big)^{\! 1/2} ~~ ,
\eea
for Dirac fermions, and 
\bea
 \Gamma ( Z' \rightarrow \phi_{\chi}^{\dagger} \phi_{\chi}) =  \frac{g_z^2 z_{\chi}^2}{192  \pi } M_{Z'}  \Big(1-4 \frac{ m_{\chi}^2}{M_{Z'}^2 } \Big)^{\! 3/2}  \;.
\eea
for complex scalars. 

The $Z'$ widths into hadrons in the $U(1)_B$ model, for  $M_{Z'}$ in the $3 - 3.7$ GeV range 
(or more precisely $ M_{K^0}^2 \ll M_{Z'}^2/4 <  M_{D^0}^2 $ so that the decays are into mesons made up of $u,d,s$ quarks, and the phase space suppression can be neglected) 
are approximately given by
\be
\Gamma ( Z'_B \rightarrow  {\rm hadrons} ) \approx   \frac{g_z^2}{48 \pi} M_{Z'} ~~ ,
\ee
while for larger $M_{Z'}$ the width increases by a factor of up to 4/3 as decay channels involving $c$ quarks open up, and above $2m_b$ by another factor of up to 5/4.
For illustration we use a benchmark set of values for the parameters: 
\be
z_{\chi} = 1 \; {\rm or } \; 3  \; , \;\;  \frac{ m_{\chi}}{M_{Z'}} = \frac{1}{4}    ~~.
\label{eq:benchmark}
\ee
The branching fractions of $Z'$ into $\chi$ particles for this set of parameters, marked with an index 0, are given in the $U(1)_B$ model by
\beqn
&& B_0 ( Z'_B \rightarrow  \psi_{\chi}  \bar   \psi_{\chi} ) \approx  ( 42\% , 87\% ) \;\; {\rm for } \;  z_{\chi} = (1 , 3) ~~,
\nonumber \\ [2mm]
&& B_0 ( Z'_B \rightarrow \phi_{\chi}^{\dagger} \phi_{\chi}) \approx ( 11\% , 52\% ) \; \; {\rm for } \;  z_{\chi} = (1 , 3)  ~~,
\eeqn
for  $M_{Z'}$ in the $3 - 3.5$ GeV range, and by somewhat smaller values as  $M_{Z'}$ increases above $2M_{D^0}$.

The $Z'$ widths into hadrons in the $U(1)_{ds}$ model, for $M_{Z'} \gtrsim 3$ GeV 
(where  the phase space suppression can be neglected for  decays  into mesons made up of $s$ or $d$ quarks) 
are given by
\be
\Gamma ( Z'_{ds} \rightarrow  {\rm hadrons} ) \approx   \frac{g_z^2}{16 \pi} M_{Z'} ~~ ,
\ee
The branching fractions of $Z'$ into $\chi$ particles for the above set of parameters in the $U(1)_{ds}$ model are
\beqn
&& B_0 ( Z'_{ds} \rightarrow  \psi_{\chi}  \bar   \psi_{\chi} ) \approx  ( 25\% , 75\% ) \; {\rm for } \;  z_{\chi} = (1 , 3)    ~~,
\nonumber \\ [2mm]
&& B_0 ( Z'_{ds} \rightarrow \phi_{\chi}^{\dagger} \phi_{\chi}) \approx ( 5.1\% , 33\% )  \; {\rm for } \;  z_{\chi} = (1 , 3) ~~.
\eeqn
for $M_{Z'} \gtrsim 3$ GeV, and decrease for smaller  $M_{Z'}$.

We now turn to deriving the constraints on the $Z'$ in the $1-10$ GeV mass range, in the $U(1)_B$ and $U(1)_{ds} $ models.
 
\bigskip

\subsection{Limits from monojet searches}

Hadron colliders set bounds on light $Z'$ via mono-jet and mono-photon searches.
For $ M_{Z'} <10 $ GeV the strongest constraint comes from the CDF search 
 $ p \bar p \rightarrow j + \slash \!\!\!\! E_T $ \cite{Aaltonen:2012jb}, and is given by \cite{Shoemaker:2011vi}:
 \be
  g_z \left[ B( Z' \rightarrow \chi \bar \chi) \right]^{1/2}  <  \left\{ 
\begin{array}{c}   \, 0.12  \; ,  \;   {\rm \;\; for }  \;\;  U(1)_B  ~~,
\\ [3mm] 
\, 0.11 \; ,  \;  {\rm \;\; for }  \;\;   U(1)_{ds}    ~~.
\end{array} \right.
\ee
The limits from ATLAS \cite{Aad:2011xw,ATLAS:2012ky} and CMS \cite{Chatrchyan:2012me,
Khachatryan:2014rra} are weaker due to stronger cuts imposed on missing energy and the jet $p_T$.
In Fig.~1 we show the regions in the $(M_{Z'},g_z)$ plane excluded by these constraints for the benchmark values Eq.~(\ref{eq:benchmark}).

\bigskip

\subsection{Invisibile quarkonium decays} 

The searches for an invisible $ \Upsilon $ decay constrain the $U(1)_B$ model, while for the $U(1)_{ds}$ model there is no such constraint since the $Z'$ does not couple to $b$ quarks.
The $Z'_B$ exchange induces an $\Upsilon \rightarrow \chi \bar\chi$ decay, with \cite{Graesser:2011vj}:
\bea
\frac{B( \Upsilon \rightarrow \text{invisible} )}{B( \Upsilon \rightarrow \mu^+ \mu^-)} =  \frac{ 4 g^4_z z_\chi^2 }{ g^4 \sin^4\!\theta_W} \left( \frac{M^2_{Z'}}{M^2_{\Upsilon}} - 1\right)^{\! -2}  \;\; .
\eea
The most stringent bound on the $\Upsilon$ invisible branching fraction has been set by the BaBar Collaboration \cite{Aubert:2009ae},
$B ( \Upsilon \rightarrow \text{invisible} ) < 3 \times 10^{-4}$
at the $ 90 \% $ confidence level. This implies that the shaded region labelled ``$ \Upsilon$" in Fig. 1 is excluded.
Similarly, a limit for $M_{Z'}$ near 3 GeV arises from 
$J/\psi$ 
decays, with the limit on invisible branching fraction given by $B ( J/\psi \rightarrow \text{invisible} ) < 7 \times 10^{-4}$ \cite{Ablikim:2007ek}.


\bigskip

\subsection{Monophoton limits from BaBar data}

At tree level the $Z'$ does not couple to leptons, but 
at one loop a kinetic mixing,
$  - (\epsilon_B/2) Z'_{\mu \nu} F^{\mu \nu},$  is generated. Therefore,  bounds from dark photon searches apply also to a leptophobic $Z'$.
  In the 1 GeV $ < M_{Z'}< 10$ GeV  mass range  the strongest constraint comes from the BaBar monophoton search reinterpreted in terms of invisibly decaying  $Z'$ produced along with a single photon in $e^+ e^-$ collisions \cite{Essig:2013vha}.
In the $U(1)_B$ model the kinetic mixing at the BaBar center of mass energy ($ E \sim 10 $ GeV)  is \cite{Graesser:2011vj}
\bea
\epsilon_{B}( 10 \;  \text{GeV}) \sim 10^{-2} g_z \;  ,
\label{babar}
\eea
 while for the $U(1)_{ds} $ model  the mixing is generated only below  the strange  quark mass and it is negligible.
In  Fig.~\ref{fig:bounds} we present the bounds on $\gamma Z'$ production taken from Fig. 5 of  \cite{Essig:2013vha} and interpreted as a bound on $ g_z$  using  Eq.~(\ref{babar}).

\bigskip

\subsection{Anomaly cancellation versus collider limits on fermions}

The inclusion of a leptophobic gauge group $U(1)_z$ requires new electrically-charged fermions which are vector-like with respect to the SM gauge group in order to cancel the gauge anomalies.
These fermions acquire a mass $m_f$ through a Yukawa coupling to a scalar $\varphi$ whose VEV breaks $U(1)_z$. 
The collider limits on $m_f$ then translate then into an upper bound on the gauge coupling  \cite{Dobrescu:2014fca}:
\be
g_z =  \frac{ \sqrt{2}  \lambda M_{Z'}}{z_\varphi m_f}  \lesssim   5.4 \times 10^{-2}\,  \frac{ 1 }{z_\varphi}  \left(\frac{M_{Z'}}{1 \; {\rm GeV}}\right)\! \left(\frac{100 \; {\rm GeV}}{m_f} \right) ~~,
\label{eq:anomaly}
\ee
where $z_\varphi$  is the $U(1)_z$ charge of $\varphi$, $\lambda$ is the Yukawa coupling, and 
we imposed a perturbativity bound $ \lambda \lesssim 3.8 $.

\begin{figure}[t]
  \begin{center}
    \includegraphics[scale=.6]{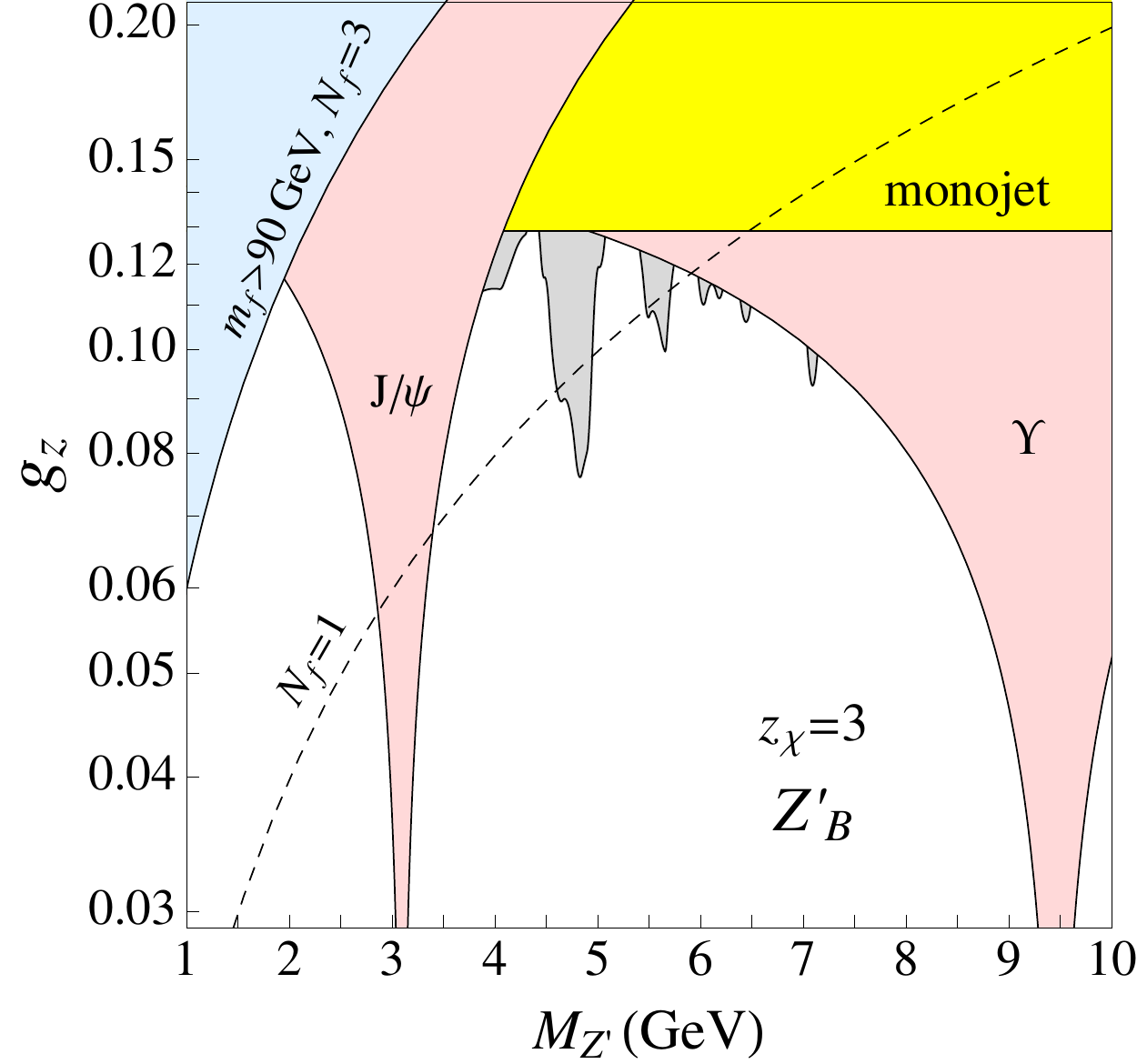} \includegraphics[scale=.6]{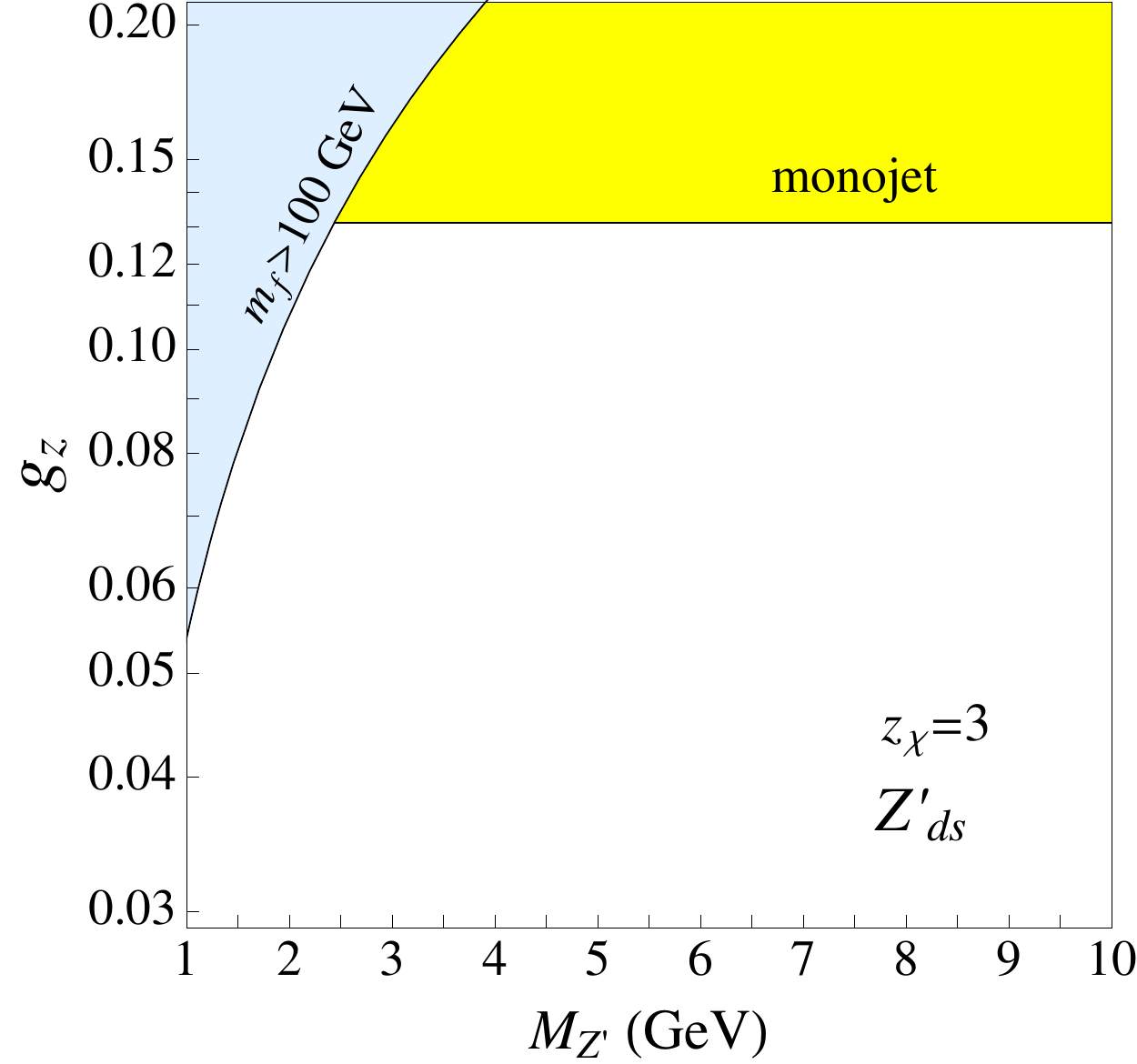}
   \caption{Constraints on the $U(1)_B$ (left panel) and $ U(1)_{ds} $ (right panel) models from monojet collider searches (upper right-hand region), collider bounds on new fermions required  to cancel gauge anomalies (upper left-hand corner), and quarkonium decays (regions labelled by $J/\psi$ and    $\Upsilon$). 
   The ragged (gray) region in the center of the left panel is due to fluctuations in the BaBar monophoton search.
   }
    \label{fig:bounds}
  \end{center}
 
\end{figure}

In the $U(1)_B$ model, $z_\varphi =3$ if the minimal set of vectorlike fermions is included.
If the charged fermions are almost degenerate with the neutral ones so that their collider signature involves only soft leptons, then 
they can be as light as $m_f = 90$ GeV, which is the LEP limit. 
If $N_f$ copies of the minimal set of vectorlike fermions are included, then $z_\varphi =3/N_f$ (see \cite{Dobrescu:2014fca} for a more detailed discussion).  
Large values of $N_f$ would increase the collider limit on $m_f$. 
The region excluded by Eq.~(\ref{eq:anomaly}), shown in the left panel of Fig.~\ref{fig:bounds}, is above the solid line labelled  ``$m_f > 90$ GeV, $N_f = 3$" in the case of three sets of vectorlike fermions,
or above the dashed line labelled ``$N_f =1$" in the minimal $U(1)_B$ model.

In the $U(1)_{ds}$ model, $z_\varphi =1$, the LEP limit on $m_f$ is about 100 GeV, and there is less flexibility in changing the fermion content.
The region excluded by Eq.~(\ref{eq:anomaly}) in the right panel of Fig.~\ref{fig:bounds} is above the line labelled ``$m_f > 100$ GeV".

\bigskip

\section{Viable DM scenarios}

Let us outline some possible scenarios which give rise in our framework  to a viable DM candidate.
Since we are interested in DM of mass $ m_{\chi}$  below a few GeV, direct detection bounds are currently very mild \cite{Angloher:2014myn,2012PhLB..711..264B, Agnese:2013jaa}. 

\begin{figure}[t]
  \begin{center}
    \includegraphics[scale=.6]{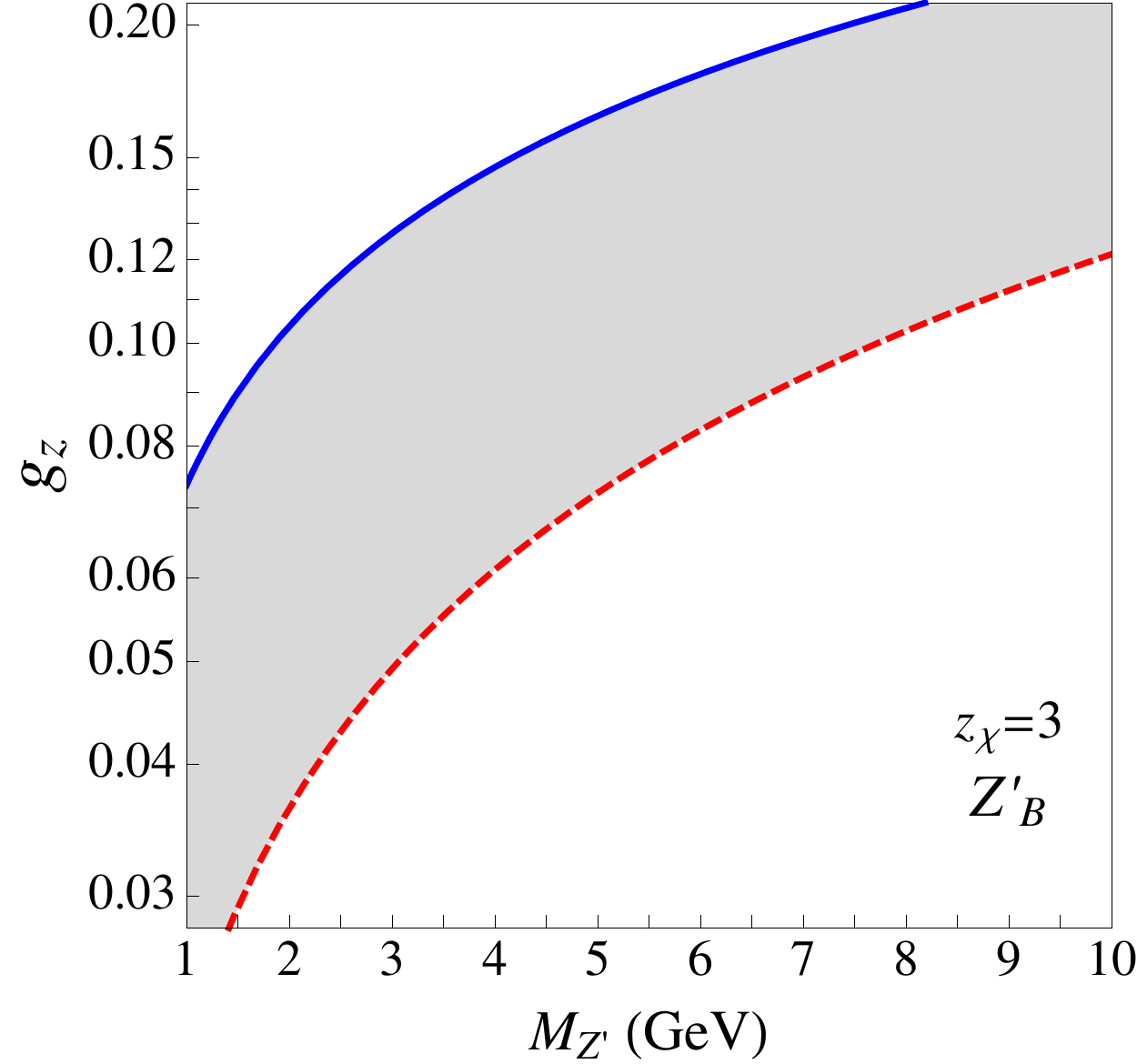} \includegraphics[scale=.6]{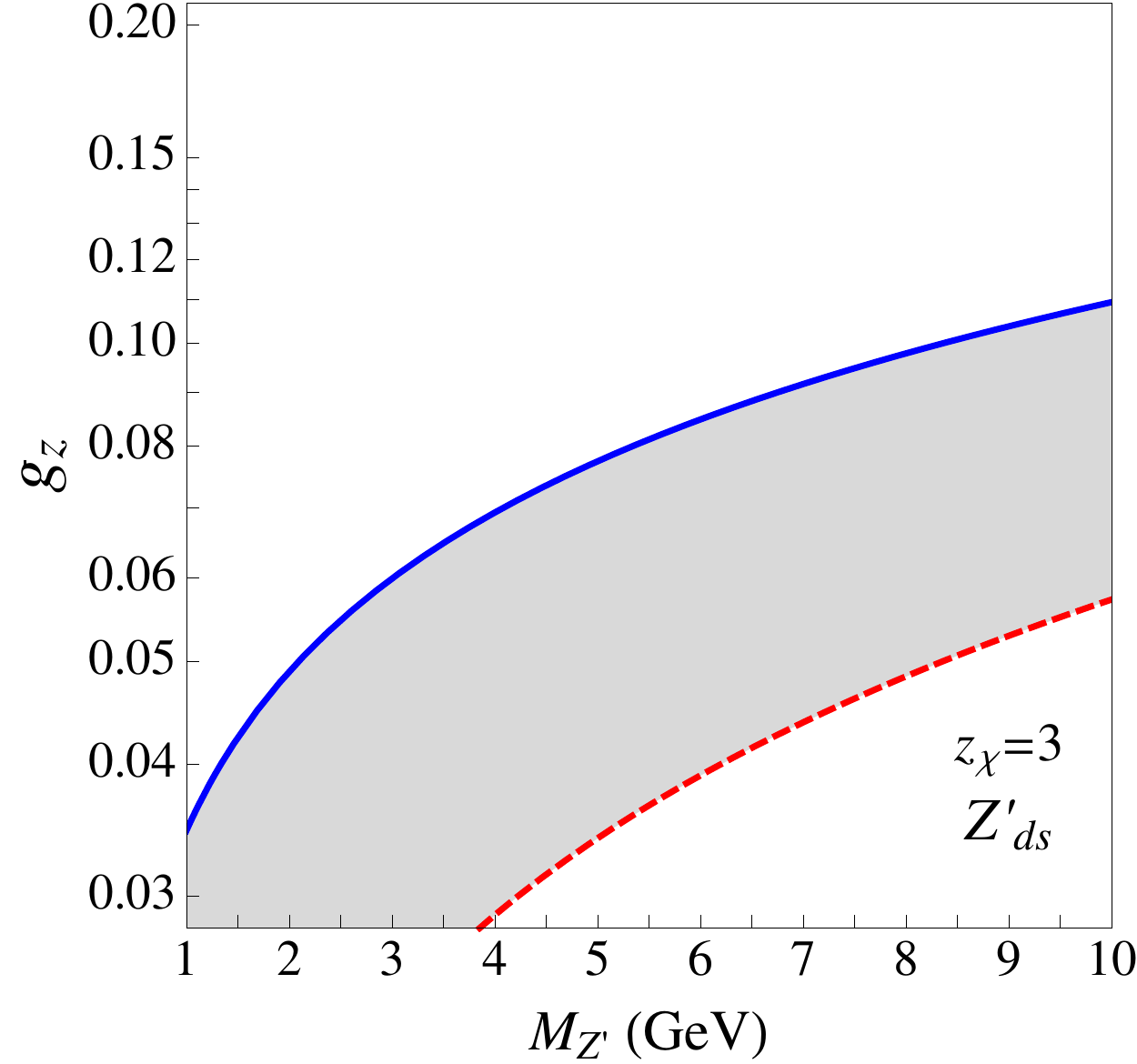}
   \caption{CMB constraints for fermonic DM $\psi_{\chi} $ in the $U(1)_B$  (left panel) and $ U(1)_{ds} $ (right panel) models. 
The region above the solid (blue) line is viable if $\psi_{\chi} $ is asymmetric DM or a subdominant DM component.
The region below the dashed (red) line is also CMB safe since the $s$-wave annihilation into quarks is small; the correct relic abundance is obtained for example via $p$-wave 
annihilation into new light scalars.
The shaded region requires a more complex hidden sector (asymmetric DM with the symmetric component depleted by annihilation into new states).}
    \label{fig:relic}
  \end{center}
\end{figure}

The most stringent constraint is provided by the bounds on energy injection  around redshifts $ z \sim 100-1000,$ coming from observations of the cosmic microwave background (CMB) \cite{Finkbeiner:2011dx,Galli:2011rz, Hutsi:2011vx,Lin:2011gj}.
This constrains the annihilation of DM into charged SM particles during recombination, and in particular rules out DM lighter than about 10 GeV  if it annihilates via  $s$-wave processes. Therefore, CMB forces the dominant annihilation to be $p$-wave suppressed or to go into neutrinos.
In our scenario a Dirac fermion $ \psi_{\chi} $  annihilates  into quarks  via  $s$-wave processes, and the thermal averaged cross section times velocity is  \cite{Lin:2011gj}
\begin{align}
\langle \sigma( \psi_\chi \bar  \psi_\chi  \rightarrow q \bar q) v\rangle_{Z'_B}= \frac{2}{9}  \langle \sigma( \psi_\chi \bar \psi_\chi  \rightarrow q \bar q) v\rangle_{Z'_{ds}}=
 \frac{z^2_{\chi}  g_z^4m_\chi^2}{ 48 \pi( M^2_{Z'}-4 m^2_\chi )^2}  \;  \;  .
 \label{swave}
\end{align}
 Therefore, the CMB bound implies that  $\psi_{\chi}$ can be a  DM particle only if it is part of a  hidden sector that is more complex than 
  the minimal model of Eq. (\ref{MinimalModel}). One possibility is to interpret the CMB bound  as an upper limit on the $s$-wave annihilation into SM particles, that is \cite{Lin:2011gj}:
\bea
\langle \sigma( \psi_\chi \bar  \psi_\chi  \rightarrow q \bar q) v\rangle  \,  \lesssim \, \frac{0.1\; \text{pb}}{f}  \left (\frac{m_\chi}{1\; \text{GeV}}\right),
\label{CMB1}
\eea
 where the ionizing efficiency factor is $ f\approx 0.2$ for pions.
 Since the annihilation is suppressed, $ \langle\sigma v\rangle   \ll  1 $ pb,  the minimal model leads to overabundant DM, and therefore needs to be extended.
 A simple extension, outlined in \cite{Batell:2014yra}, includes a scalar  $ \eta$ that has a Yukawa coupling,  $ y_1 \eta \bar \psi_{\chi}  \psi_{\chi} ;$ if $m_{\eta} < m_{\psi_{\chi}} <M_{Z'}/2 $ the annihilation $ \psi_{\chi} \bar  \psi_{\chi} \rightarrow 2 \eta $ dominates, and gives the correct relic abundance, $\it{e.g}$  for  $ y_1 \sim 0.05$, $m_{\psi_{\chi}} =1$ GeV and
 $m_{\eta} =100$ MeV. This annihilation mode is  $p$-wave suppressed and therefore CMB safe.
 The $\eta $ scalar can then decay into SM particles via a small  Higgs portal coupling.
The  condition in Eq.(\ref{CMB1}) is satisfied for  values of  $ g_z $  below the dashed red curves in  Fig. \ref{fig:relic}. 

Another way to satisfy the CMB constraint  is to suppress the $\psi_{\chi}$ relic density rather than the annihilation cross section into SM charged particles.
This possibility requires a substantially larger cross section than the one corresponding to the correct relic abundance \cite{Lin:2011gj}:
\bea
\langle \sigma( \psi \bar  \psi  \rightarrow q \bar q) v\rangle   \gtrsim 16\;  \text{pb}.
\label{CMB2}
\eea
This is self-consistent if $\psi_{\chi} $ is a subdominant DM component. Alternatively the minimal model can be part of an asymmetric DM model \cite{Zurek:2013wia}, so that  Eq.(\ref{CMB2}) represents the condition of depletion of the symmetric component \cite{Lin:2011gj}. 
Indeed in models of asymmetric DM, annihilation during recombination can be easily  suppressed.
The region above the solid blue lines in  Fig. \ref{fig:relic} satisfies Eq.~(\ref{CMB2})  for $ m_\chi= M_{Z'}/4$.

The shaded region in
Fig. \ref{fig:relic},  between the two lines corresponding to the two scenarios just described, is not necessarily ruled out: if  both ingredients discussed above are present then the CMB constraints can be accommodated. 
We see that the CMB constraints are very model dependent, and hence there is still a large region of the parameter space yet unexplored which leads to robust DM  scenarios where Dirac fermions interact with SM quarks via $s$-wave processes.

In the case of scalar DM, $\phi_{\chi} $, the  annihilation cross section into quarks is  \cite{Boehm:2003hm}:
\begin{align}
\langle  \sigma( \phi_\chi \phi_\chi^*\rightarrow q \bar q) v\rangle _{Z'_B}= \frac{2}{9}  \langle \sigma( \phi_\chi \phi_\chi^* \rightarrow q \bar q) v \rangle _{Z'_{ds}}=
 \frac{v^2  z^2_{\chi}   g_z^4 m_\chi^2}{288  \pi (M^2_{Z'}-4 m^2_\chi )^2} \;  \;  ,
 \end{align}
where $v \sim 0.3 $ is the DM velocity at freeze out. This is $p$-wave suppressed, and hence CMB safe.
Large gauge couplings are typically required in order to achieve the correct relic abundance (which requires   at freeze out  $ \langle\sigma v\rangle \sim 1.5$ pb  for light dark matter \cite{Steigman:2012nb}) and these could then  be possibly already excluded by the current constraints.
For $m_\chi  \approx M_{Z'}/4$ and $z_{\chi}=3$ the correct  relic abundance is obtained with $ g_z \sim 0.06  \, (M_{Z'}/1 \text{GeV})^{1/2}$ for the $U(1)_B$ model and $ g_z \sim 0.04 \,  (M_{Z'}/1 \text{GeV})^{1/2}$ for the $U(1)_{ds} $ model.
Since the present bounds on scalar DM are similar to the ones for fermion DM presented in Fig. \ref{fig:bounds},
 we conclude that 
there are still open regions of the parameter space where the minimal model Eq.~(\ref{MinimalModel}) gives a scalar thermal DM candidate.

\section{DM production through proton scattering off nucleons}
\label{sec:}\setcounter{equation}{0}

Having examined the bounds on GeV-scale leptophobic $Z'$ bosons decaying into DM particles, we now proceed to discuss the potential sensitivity of proton fixed target experiments to this scenario.

We assume that the $Z'$ boson is produced on-shell and then decays into DM particles ($ M_{Z'}> 2 m_{\chi})$. This way the DM particles are produced resonantly: $ p N  \rightarrow Z' \rightarrow \chi \bar \chi $, where $N$ indicates the nucleon  inside the target.
The cross section for proton-nucleon scattering,  computed within the parton model, is 
\bea 
\sigma (p N  \rightarrow  \chi \bar \chi) = \int{ dx_1 dx_2 \sum_q{ f_{q|N}(x_1) f_{\bar q|N}(x_2) \hat \sigma_q(x_1 x_2 s)}} \, B(Z'\rightarrow \chi \bar \chi) ~~,
\eea
where $ B(Z'\rightarrow \chi \bar \chi)$ is the branching fraction of the $Z'$ boson into DM particles.
If the vector boson is  produced on-shell, the tree-level partonic cross section is
\bea
\hat \sigma_q(\hat s)= \frac{ g_z^2}{3} \left(z_{q_L}^2 + z_{q_R}^2 \right) \delta(\hat s -M^2_{Z'} )   ~~.
\eea
In the $U(1)_B $ model $z_{q_L} = z_{q_R}  =1/3$, so that the proton and the neutron cross sections are the same.
In the $U(1)_{ds}$ model only the right-handed $d$ and $s$ quarks have nonzero charges ($z_{q_R}^2=1$ for $q=d,s$),
leading to different proton-neutron and proton-proton cross sections:  $ \sigma ( p n \rightarrow Z_{ds}' ) \simeq 2  \sigma ( p  p \rightarrow Z_{ds}' ) $.
As a result, the average proton-nucleon cross section is material dependent. 
For a target of atomic  mass  $A_{\rm T}$  and atomic number $Z_{\rm T}$, the average $pN$ cross section is 
\bea
\sigma (pN \to\chi\bar\chi)_{\rm T} \simeq  \frac{1}{A_{\rm T} }  \; \Big(Z_{\rm T} \;  \sigma ( p p \rightarrow  \chi \bar \chi)+(A_{\rm T}-Z_{\rm T}) \; \sigma ( p n \rightarrow  \chi \bar \chi) \Big) ~~.
\label{Prod}
\eea

Comparing the $Z'$ production rate with the total proton-proton cross section, $ \sigma (pp)$, which 
for a 120 GeV beam is given by $ \sigma (pp) \approx 40$ mb  (see fig. 46.10 \cite{Agashe:2014kda}), we find the number of DM particles produced in the target: 
\bea
N_{\chi }^{\rm T} =   \; \frac{2 N_{{\rm POT} }}{\sigma (p p) }  \; \sigma (pN \to\chi\bar\chi)_{\rm T} \;  ,
\label{Prod}
\eea
where $N_{{\rm POT} } $ is the number of protons on target.

\begin{figure}[tp]
  \begin{center}\hspace*{-0.5cm}
   \includegraphics[scale=.7]{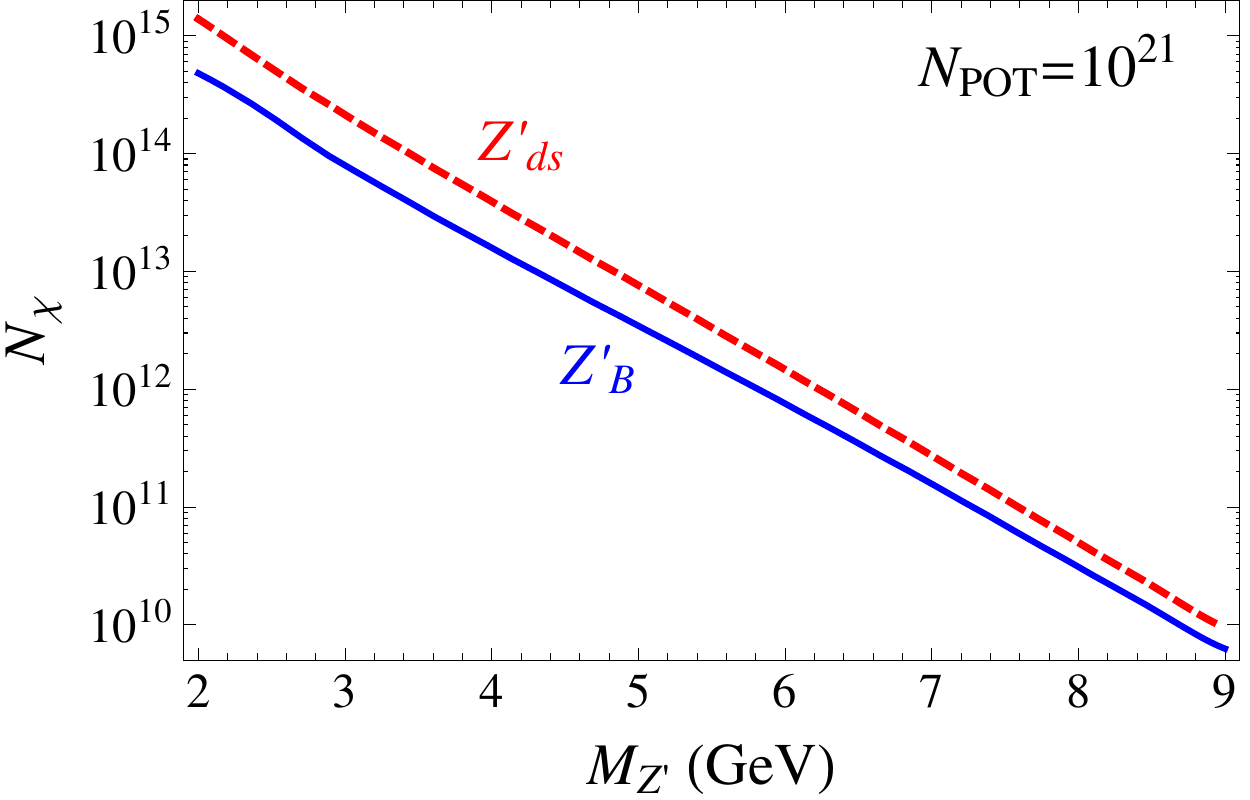}
      \caption{ Number of $\chi$ or $\bar\chi$ DM particles produced 
      for $N_{\rm POT} = 10^{21}$  protons  of 120 GeV scattering off a  fixed target which is thick enough to stop all incoming protons. The two lines 
      are predicted in the $U(1)_B$ (solid blue line) and $U(1)_{ds}$ models (dashed red line, for $A_T = 2 Z_T$) 
      with  $g_z=0.1$. The $Z'\to \chi \bar\chi$ branching fraction used here is 100\%, corresponding to  $z_\chi \gg 1$; for smaller $z_\chi$ the branching fraction
      depends on $M_Z'$, $m_\chi$ and the $\chi$ spin (see Section 2).}
       \label{fig:XS}
     \end{center}
\end{figure}

Using  MadGraph 5 \cite{Alwall:2011uj} to compute the production cross section, and FeynRules \cite{Christensen:2008py}
to implement the $Z'$ models, we find the number $N_{\chi }^{\rm T} $  of produced DM particles shown in Fig. \ref{fig:XS} for 
$10^{21}$ protons on target.
We focus on $M_{Z'} > 2$ GeV because the validity of the parton model is questionable in the case of lighter $Z'$ production.
The $Z'_{ds}$ line shown in Fig. \ref{fig:XS}  corresponds to an isospin-symmetric target ($A_T = 2Z_T$).
More generally, the number of DM particles produced has only a mild dependence on $Z_T/A_T$ (and is material independent in the $Z'_B$ model).

The value of the gauge coupling used in Fig. \ref{fig:XS} is $g_z= 0.1$; for other values, $N_\chi$ scales as $(g_z/0.1)^2$.
The branching fraction for $Z'\to \chi \bar\chi$ used in Fig. \ref{fig:XS} is 100\%; more realistic choices, discussed in Section 2, depend on $m_\chi$, $z_\chi$
and on whether $\chi$ is a fermion or a scalar
(in the case where $\chi$ is a Dirac fermion and $z_\chi = 3$, the branching fraction is large, of about 87\% for $Z'_B$ and 75\% for $Z'_{ds}$). 


\bigskip

\section{DM flux through detectors}
\label{sec:}\setcounter{equation}{0}

We now proceed to compute how many of the produced dark matter particles pass through detectors, as well as their energy distribution. 
We will discuss both off-axis and on-axis detectors, with examples given by the NO$\nu$A  and MINOS near detectors. 

\begin{figure}[b]
  \begin{center}
    \includegraphics[scale=1.]{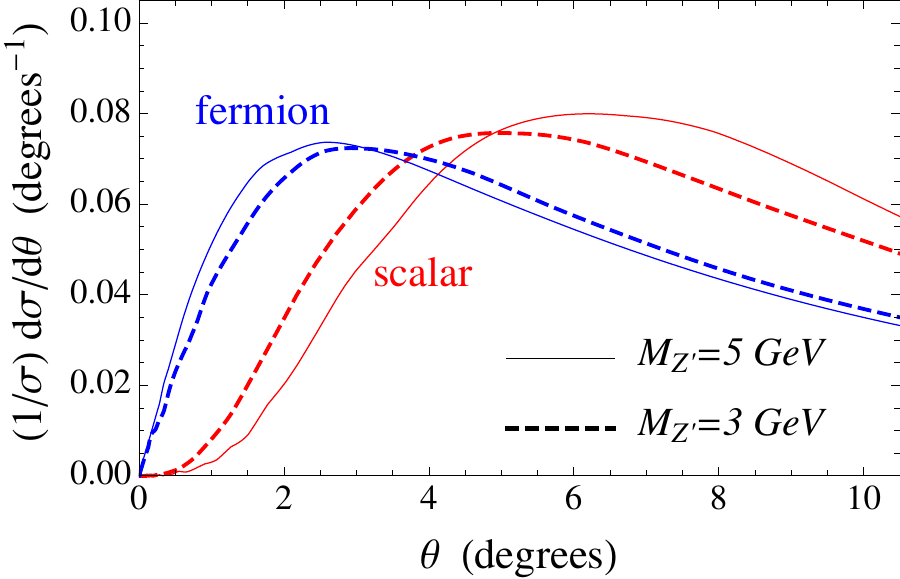}
    \caption{Polar angle distribution in the lab frame of the DM particles produced in the $pN \to Z' \to \chi\bar \chi$ process, for $M_{Z'} = 3$ GeV (dashed lines) or $ M_{Z'} = 5$ GeV (solid lines), 
    when $\chi$  is a Dirac fermion or a complex scalar. 
    }
 \label{ba}
  \end{center}
\end{figure}

\subsection{Angular distribution of DM particles}

Let us denote  the polar  angle in the lab frame ({\it i.e.,} the angle between the direction of one of the DM particles 
produced in $Z'$ decays and the beam direction) by $\theta$.
Using the output of the MadGraph simulation, 
we obtain the polar angular distributions shown in Fig.~\ref{ba} in the cases where the DM particle is a Dirac fermion or a complex scalar.   
The fermion angular distribution is more spread and peaked towards smaller angles than the scalar one. This different behavior can be understood by considering the scalar and fermion distributions in the center-of-mass frame,  $d\sigma/d\theta \propto  (1 \mp \cos^2{\theta})\sin\theta$, and then boosting to the lab frame. 
Note that our choice of vector coupling of $Z'$ to the fermion DM implies that the angular distribution is the same independently of the $Z'$ couplings to quarks.

The Carbon target used in the NuMi beam line stops about  $85 \% $ of the incoming protons, while  the remaining $15 \%$ of protons travel through the 675 m long vacuum pipe and 
hit the absorber, which is mostly made of iron. 
These two sets of protons give comparable contributions to the total number of DM particles inside the MINOS and NO$\nu$A near detectors. The smaller number of protons 
reaching the absorber is compensated by the larger coverage of the detector along the azimuthal angle $\phi$. 
In Table I we list the approximate angular cuts required for a DM particle produced in the absorber or the target to reach 
the MINOS or NO$\nu$A near detectors.

\begin{table}[t]
\renewcommand{\arraystretch}{1.5}
\begin{center}
\begin{tabular}{|c|c||c|ccc|cc|}\hline 
Detector & DM source & distance & $\theta_{\rm min}$ & $\theta_{\rm max}$  &  $\phi_{\rm max}$  & $\epsilon_{\rm det} $(fermion)  & $\epsilon_{\rm det} $(scalar) \\  [-0.05em] \hline
MINOS & absorber  & 270 m  & 0        & $0.48^\circ $   
 & $180^\circ $ &  $6 \times 10^{-3} $ &  $10^{-4} $  \\ [0.1em]  
MINOS &  target      & 950 m  &  0   & $0.19^\circ $     &    $180^\circ $    &   $8 \times  10^{-4} $ &  $3 \times 10^{-6} $ \\ [0.1em]  \hline
NO$\nu$A  & absorber & 240 m & $2.6^\circ $  
& $3.6^\circ $ 
& $18^\circ $ 
& $3 \times 10^{-3} $  &  $2 \times 10^{-3} $ \\ [0.1em]  
NO$\nu$A  &  target  & 920 m &  $0.68^\circ $   & $0.93^\circ $         & $18^\circ $      &  $4 \times 10^{-4} $  &  $3 \times 10^{-5} $ \\   \hline
\end{tabular}
\medskip \\
\caption{\small Geometrical parameters for particles produced in the absorber or the target and passing through the MINOS or NO$\nu$A near detectors. The polar angle 
satisfies $\theta_{\rm min} < \theta < \theta_{\rm max}$, while the azimuthal angle satisfies $0 \leq \phi \leq \phi_{\rm max}$. The geometric acceptance of the detector $\epsilon_{\rm det} $ (shown here for $M_{Z'} = 3$ GeV)
depends on the DM spin.
}
\label{table:Detect}
\end{center}
\end{table}

We compute the geometrical acceptance of the detector, $\epsilon_{\rm det} $,  by imposing angular cuts 
on the DM particles produced in the simulated events.
In the case of a Dirac fermion,  we find $\epsilon_{\rm det} \gtrsim O(10^{-3})$   
both for MINOS and NO$\nu$A near detectors for particles produced at the absorber, while for particles produced at the target the acceptance is smaller by an order of magnitude 
due to the larger distance. The values of the acceptance are given in Table I for $M_{Z'} = 3$ GeV.
For a scalar, $d\sigma/d\theta$ vanishes faster for $\theta \to 0$, so that the 
 acceptance of on-axis detectors is suppressed. 
Therefore, this offers a possibility to measure the spin of a discovered DM particle via a parallel MINOS and NO$\nu$A analysis.

Higher-order processes that include real radiation, $ p p \rightarrow Z' + \rm{jets}$, can potentially change the scalar angular distribution. 
As a crude approximation, we computed the tree-level production of $Z'$  together with one or two hard jets, imposing a jet-$p_t$ cut of 1 GeV, and we found that these processes 
are not large enough compared to $ p p \rightarrow Z'$ to change qualitatively the above result. \\

\subsection{Energy distribution of DM particles}

 \begin{figure}[b]
  \begin{center}  \includegraphics[scale=0.72]{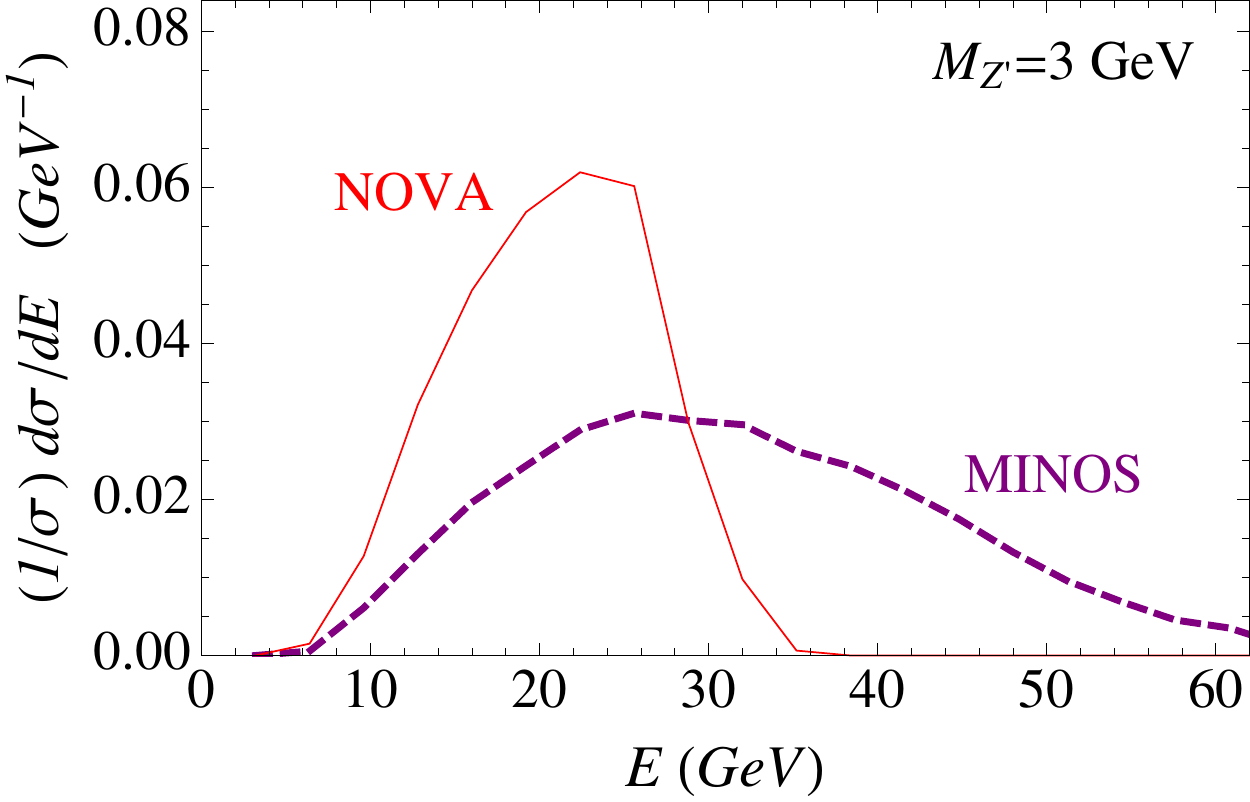}
   \caption{Energy distribution of fermonic DM particles produced in the absorber and passing through the NO$\nu$A or MINOS  near detectors for  $ M_{Z'}= 3$ GeV. }
    \label{energyDM}
  \end{center}
\end{figure}

The energy distributions of DM particles that enter the NO$\nu$A and MINOS  near detectors are shown in Fig.~\ref{energyDM} 
for $ M_{Z'} =3$ GeV. 
The DM particles inside the detectors are fairly energetic compared to the neutrinos (produced mostly in pion decays), especially for NO$\nu$A.
The neutrinos enter the NO$\nu$A near detector with a peak energy around 3 GeV; a tail of high-energy neutrinos  ($E_\nu \gtrsim 10$ GeV) 
is produced mostly by kaon and heavier meson decays. 
The difference  in the energy profile  between DM and neutrino can be used to reduce the neutrino background.
This can be done considering processes where DM transfer a significant energy to nucleus, and imposing a cut on the energy of the hadronic activity.

The difference between energy distributions of DM and neutrinos is less pronounced for on-axis detectors such as MINOS, where the neutrinos have a wider energy distribution with a long tail \cite{Adamson:2011ku,MINOS:1998nfa,Thron:1996vc}.
However, a dedicated search for MINOS near detector would also be possible and highly interesting, particularly for reasons mentioned in section 5.1 related to the distinction between
DM beams of scalars or fermions.

\section{DM  scattering inside the detector} 
\label{sec:}\setcounter{equation}{0}

DM particles may scatter off the nucleons in the detector by exchanging a $Z'$ boson in the $t$ channel, and producing neutral-current events. 
Let us study the interaction between the DM particles and nucleons inside  the detector. 

The cross section for DM interacting with nucleons 
can be much larger than the neutrino one since it is 
mediated by a lighter boson. For momentum transfers much smaller than $M_{Z'}$, the ratio of the DM to neutrino
cross sections for neutral-current events is 
\bea \hspace*{-1.cm}
\frac{\sigma( \chi N \rightarrow \chi j)}{\sigma( \nu N \rightarrow \nu j)}  & \approx & C_N (Z')\, \frac{z_\chi^2}{4} \, \left( \frac{g_z}{g} \cos\theta_W \frac{M_Z}{M_{Z'}} \right)^{\! 4}
\label{eq:sigmaratio}
\\ [2mm]
& \approx & 23 \, C_N (Z')  \, z_\chi^2\,  \left( \frac{g_z}{0.1}\right)^{\! 4}
\left( \frac{ \text{4 GeV}}{ M_{Z'}} \right)^{\! 4}  ~~,   \nonumber 
\eea
where $j$ stands for any hadronic final state.
$C_N (Z')$ is a coefficient of order one that depends on whether the nucleon $N$ is a proton ($N=p$) or neutron ($N=n$), as well as on the 
$Z'$ model; neglecting the interactions of sea quarks and nuclear form factors, this coefficient takes the values
\bea
&& C_p (Z'_B) \approx \frac{2}{3} \left( \frac{3}{4} - \frac{5}{3} s_W^2 + 2 s_W^4 \right)^{\! -1}  \approx 1.42 \;\;\; ,  \;\;\; C_p (Z'_{ds} ) \approx  \frac{3}{2}  \, C_p (Z'_B)  \;\;  , \nonumber \\ [2mm]
&& C_n (Z'_B) \approx \frac{2}{3} \left( \frac{3}{4} - \frac{4}{3} s_W^2 + \frac{16}{9}s_W^4 \right)^{\! -1} \approx 1.25  \;\;\; ,  \;\;\; C_n (Z'_{ds} ) \approx  3 \,  C_n (Z'_B)  \;\;   ,  
\eea
where $s_W\equiv \sin\theta_W$ is evaluated at a scale of a few GeV ($\sin^2\theta_W \approx 0.235$).
For a DM charge under the new $U(1)$ group of $z_\chi = 3$, the values of the $Z'$ gauge coupling and mass shown in Fig.~1 allow
the ratio in Eq.~(\ref{eq:sigmaratio}) to be as large as $10^3$. However, even with such a large cross section for DM-nucleon scattering, the total number of DM events 
in the detector is much smaller than that of neutrino neutral-current events, because of the very large QCD production of pions and other mesons, whose decays generate the neutrino beam.


Before discussing selection cuts that reduce the neutrino background, let us compute the total number of DM events in the detector.
The average DM-nucleon scattering cross section in the detector can be written as
\bea
  \sigma ( \chi N \rightarrow \chi j)_{\rm d} = \frac{1}{A_{\rm d} } \; \Big(Z_{\rm d} \;  \sigma ( \chi  p)+(A_{\rm d}-Z_{\rm d}) \; \sigma ( \chi  n) \Big) ~,
\label{sigma-detector}
\eea
where $j$ stands for any set of hadrons.
The DM-proton and DM-neutron cross sections, $\sigma ( \chi  p)$ and $\sigma ( \chi  n)$, are functions of the incoming DM energy $E_\chi$.  
The number of DM particles which are produced in the target and enter the detector is $f_{\rm T} \epsilon_{\rm det}^{\rm T}  \, N_{\chi }$, where $f_{\rm T}$ is the fraction of incoming protons stopped in the target ($f_{\rm T} \approx 0.85$ for the NuMI beam line), 
$N_\chi$ is shown in Fig.~\ref{fig:XS}, and the geometric acceptance $ \epsilon_{\rm det}^{\rm T} $ is given in Table I.
Multiplying this number of particles by the fraction of those that have energy between $E_\chi$ and $E_\chi + d E_\chi $ (shown in Fig.~\ref{energyDM}) gives
\begin{equation}
dN_{\rm T}  (E_\chi)  =  f_{\rm T} \, \epsilon_{\rm det}^{\rm T}  \, N_{\chi } \,  \left(\frac{1}{\sigma} \frac{d\sigma}{ dE_\chi}\right)\!\! \left(pN\to \chi\bar{\chi}\right)_{\rm T} \; dE_\chi   ~~.
\label{fractionE}
\end{equation}
For a detector of density $\rho_d$ and length $L_d$, 
the number of signal events 
due to the beam produced in the target is given by
\begin{equation}
S_{\rm T}  = \frac{  L_d \; \rho_d}{m_p} \;  \int
dN_{\rm T}  (E_\chi)   \, \sigma( \chi N \rightarrow \chi j)_{\rm d}  ~~.
\label{DIS1}
\end{equation}
For the NO$\nu$A near detector $\rho_d \approx 1263$ kg/m$^3$ and $L_d \approx 14.3$ m, while for the MINOS near detector $\rho_d \approx 3237$ kg/m$^3$ and $L_d \approx 16.6$ m;
$A_d \approx 2 Z_d$ is  a good approximation for both detectors.

An  expression analogous to Eq.~(\ref{DIS1}) 
can be obtained for the number ($S_{\rm A}$) of  signal events due to the beam produced in the absorber, by replacing the quantities carrying a T index with the ones corresponding to 
the absorber (marked by an A index). Given that all incoming protons are stopped in the absorber, $f_{\rm A} = 1 - f_{\rm T}$.  The total number of signal events is thus $S_{\rm T}  + S_{\rm A}$. 

Without imposing cuts there are $O(10^7)$ neutral-current neutrino scattering events, which is a too large background for allowing sensitivity to our signal.
Thus, it is necessary to find some selection cuts that reduce the neutrino background without reducing the DM signal too much.
If we label the incoming and outgoing $\chi$ four-momenta by $\ k^{ \mu} $ and  $\ k^{\prime \mu} $\
respectively, and the momentum transfer by 
$q^{ \mu}=k^{ \mu}-k^{\prime \mu}, $\ then the DIS regime is realized for $Q^2 > m_{p}^2 $.
For lighter mediators, $ M_{Z'}<$ 1 GeV,  the most relevant region for fixed target experiments is  $Q^2 = -q^2 <$ 1 GeV \cite{Soper:2014ska} . 
This is explained by  the $Q^2$ dependence of the cross-section: $ 1/(M_{Z'}^2+Q^2)$.
For heavier $Z'$ bosons, of mass around a few GeV, we expect the DIS regime to dominate. Consequently, we expect that it is helpful to impose a cut on the energy $E_j$ 
of the hadronic activity produced by the DM particle in the detector.

The peak energy of the neutrinos that enter the NO$\nu$A detector is near 3 GeV, while the energy hadronic activity due to the neutrinos peaks at  smaller values.
We impose a cut $E_j > 2$ GeV, as stronger cuts reduce the signal too much in some cases.
We expect this cut to not be sufficient to reduce enough the huge neutrino background. Therefore, additional strategies may be required, such as  timing the delay of DM, or running in the proton beam-dump mode \cite{Batell:2009di,deNiverville:2011it,deNiverville:2012ij, Dharmapalan:2012xp,Batell:2014yra}.

  \begin{figure}[tp]
  \begin{center}\hspace*{-0.2cm}
   \includegraphics[scale=.58]{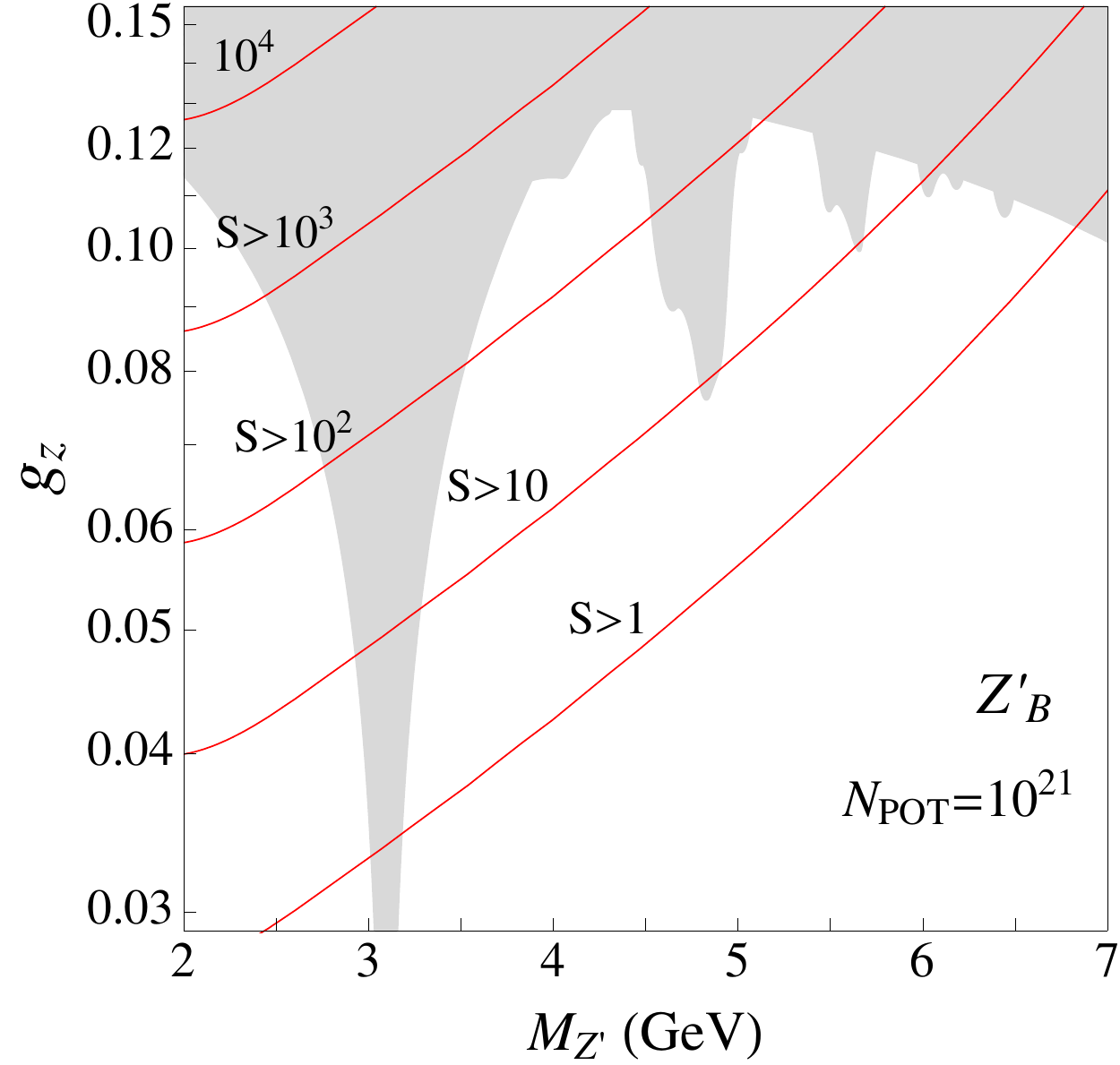} \hspace*{0.2cm}\includegraphics[scale=.58]{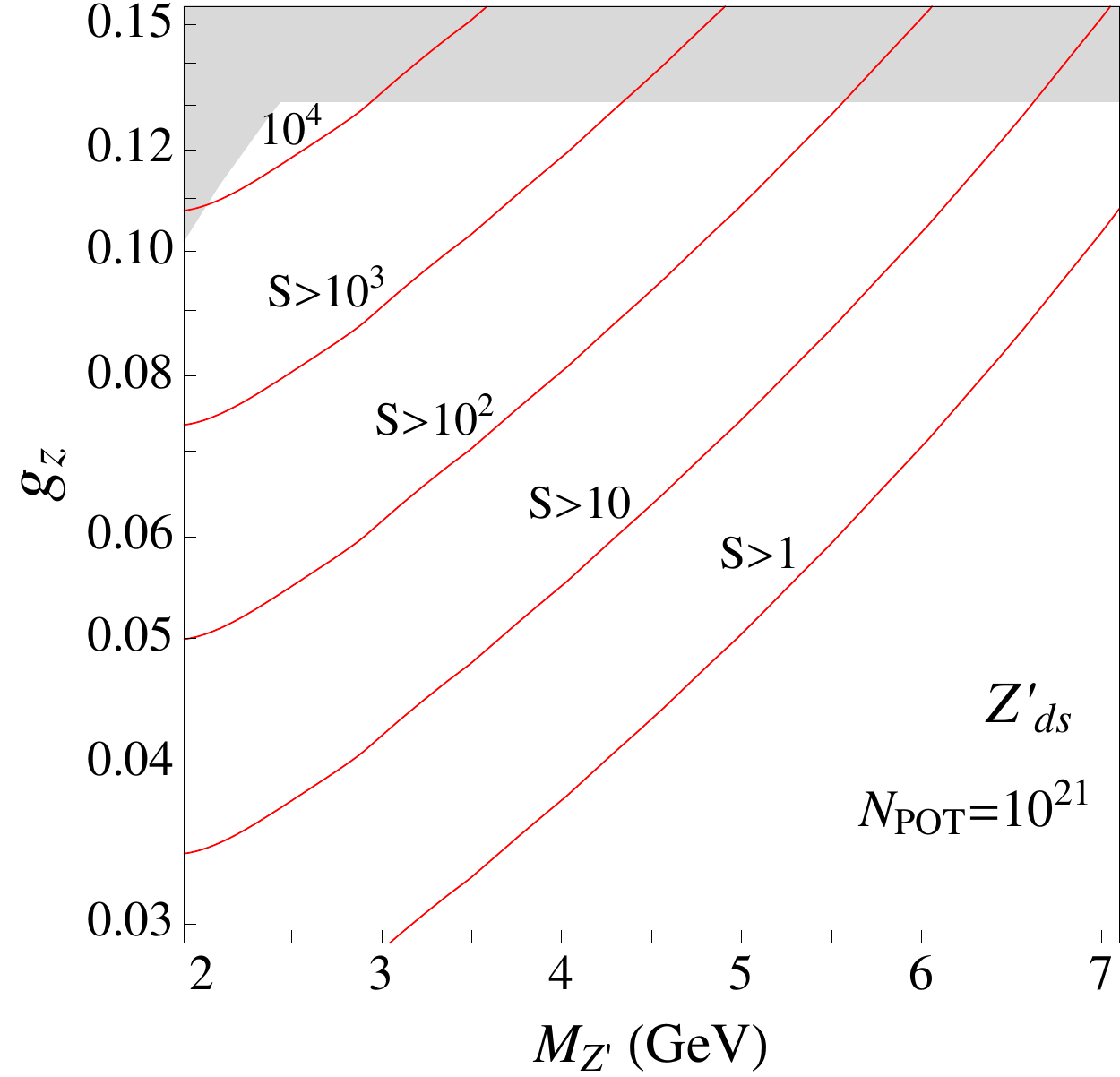}
   \caption{Predicted number of DM scattering events with hadronic energy above 2 GeV  in the  NO$\nu$A near detector, 
   shown as red contour lines, for the $U(1)_B$ (left)  and $U(1)_{ds} $ (right) models with $z_\chi = 3$. The shaded regions are excluded by other experiments (see Fig.~1).}
  \label{fig:Signal}
  \end{center}
 \end{figure}

We computed the $\sigma( \chi N \rightarrow \chi j)_{\rm d}$  cross section after this cut with MadGraph 5;
in Fig.~\ref{fig:Signal} we show the regions  in the $(M_{Z'},g_z)$ plane (above the red lines) where the number of DM scattering events in the NO$\nu$A detector  satisfies $S > 10^n$, $n = 1,...,4$, for $z_\chi = 3$.
We also show there  the regions (shaded) excluded by various experiments 
discussed in Section 2. We find that the number of signal events $S$ can be as large as $10^3$ for $Z'_B$ and $10^4$ for $Z'_{ds}$, 
with $N_{\rm POT} = 10^{21}$. 
It appears that NO$\nu$A may be able to explore a viable region of parameter space. 

The MINOS near detector may also be able to probe the case of fermionic $\chi$, 
 even though the background is larger because the neutrinos traveling closer to the axis are typically more energetic  \cite{Adamson:2011ku}.
Therefore, we urge both the NO$\nu$A  and MINOS Collaborations to perform dedicate analyses to probe the presence of a DM beam.

\bigskip

\section{Outlook}
\label{sec:}\setcounter{equation}{0}

It is important for DM searches to be as broad as possible in order to cover the wide range of allowed masses and the many potential portals to the visible sector.
Proton fixed target experiments offer the possibility to explore a region of the parameter space left unconstrained by the existing searches involving direct detection, 
collider searches, or invisible quarkonium decays.
The high beam intensity of these experiments makes them a promising ground for testing models with light DM particles.

In this paper we have studied the potential sensitivity of the neutrino near detectors to a DM beam produced at the Fermilab Main Injector, in the NuMI beam line. 
We have considered  DM candidates, either a fermion or a scalar, charged under a new leptophobic  gauge group, with the
associated $Z'$ boson having  a mass $M_Z'$ in the $1 -10$ GeV range. 
Assuming that the DM mass satisfies $m_\chi < M_{Z'}/2$, pairs of DM particles may be 
resonantly produced when the 120 GeV proton beam scatters off the target or is dumped into the absorber.

We  have found a potentially interesting reach for these experiments provided that an efficient way to reduce the neutrino background is found.
We have outlined possible solutions in this sense, focusing  on the  NO$\nu$A near detector since its off-axis position is better suited than the MINOS one for reducing the neutrino background.

Even though the NO$\nu$A near detector is better positioned for our purpose, 
the MINOS near detector can also be sensitive to a DM beam, especially in the case of fermionic DM particles.
A parallel analysis done by the MINOS and  NO$\nu$A collaborations can reveal the spin of the particle potentially discovered.
The MIner$\nu$a detector is on-axis, but about 10\% closer to the absorber than MINOS, so it may provide further tests.

The neutrino detectors along the Booster beam line  at Fermilab may also be used to probe the presence of DM beams generated at the NuMI beam line. 
The MiniBoone detector  is $6.3^\circ$ off-axis when viewed from the NuMI target \cite{Adamson:2008qj}, and at more than $90^\circ$ off-axis when viewed from the 
NuMI absorber; it also happens to be closer to the absorber by a factor of 2 compared to the MINOS near detector.
The MicroBoone detector \cite{Katori:2011uq}
is $\sim 8^\circ$ off-axis when viewed from the NuMI target, and the proposed LAr1-ND detector \cite{Admas:2013xka} would be $\sim 30^\circ$ off-axis.

The proposed LBNF \cite{Adams:2013qkq} beam line at Fermilab would have a substantially larger $N_{\rm POT}$ than the NuMI beam line. Thus, 
if a near detector is built close enough to the LBNF target or absorber, then the expected number of DM 
events can be an order of magnitude higher than in NO$\nu$A. Furthermore, the steep downwards slope of the 
proposed  LBNF beam, in conjuction 
with the shape of the DM beam (a conic shell originating at the target and another one 
originating at the absorber), offers interesting physics opportunities if two or more near detectors are placed off-axis. 

%

\acknowledgments{ 
We thank Prateek Agrawal, Brian Batell, Pilar Coloma, Patrick Fox, Martin Frank, Lisa Goodenough, Roni Harnik, Lauren Hsu, Siva Kasetti, Ian Lewis, Tongyan Lin,  David McKeen, Denis Perevalov, Robert Plunkett, Adam Ritz and David Schmitz for valuable discussions and comments.
Fermilab is operated by Fermi Research Alliance, LLC  under Contract No. DE-AC02-07CH11359 with the United States Department of Energy.
}

\bibliography{DMneut}

\providecommand{\href}[2]{#2}\begingroup\raggedright\begin{thebibliography}{10}

\bibitem{Agashe:2014kda}
{\bf Particle Data Group} Collaboration, K.~Olive {\em et.~al.}, {\it {Review
  of Particle Physics}},  {\em Chin.Phys.} {\bf C38} (2014) 090001.

\bibitem{Cushman:2013zza}
P.~Cushman, C.~Galbiati, D.~McKinsey, H.~Robertson, T.~Tait, {\em et.~al.},
  {\it {Working Group Report: WIMP Dark Matter Direct Detection}},
  \href{http://xxx.lanl.gov/abs/1310.8327}{{\tt arXiv:1310.8327}}.

\bibitem{Dobrescu:2014fca}
B.~A. Dobrescu and C.~Frugiuele, {\it {Hidden GeV-scale interactions of
  quarks}},  {\em Phys.Rev.Lett.} {\bf 113} (2014) 061801,
  [\href{http://xxx.lanl.gov/abs/1404.3947}{{\tt arXiv:1404.3947}}].

\bibitem{Tulin:2014tya}
S.~Tulin, {\it {New weakly-coupled forces hidden in low-energy QCD}},  {\em
  Phys.Rev.} {\bf D89} (2014) 114008,
  [\href{http://xxx.lanl.gov/abs/1404.4370}{{\tt arXiv:1404.4370}}].

\bibitem{Anderson:1998zza}
K.~Anderson, B.~Bernstein, D.~Boehnlein, K.~R. Bourkland, S.~Childress, {\em
  et.~al.}, {\it {The NuMI Facility Technical Design Report}},  {\em \rm report
  FERMILAB-DESIGN-1998-01} (1998).

\bibitem{Patterson:2012zs}
{\bf NOvA} Collaboration, R.~Patterson, {\it {The NOvA Experiment: Status and
  Outlook}},  {\em Nucl.Phys.Proc.Suppl.} {\bf 235-236} (2013) 151--157,
  [\href{http://xxx.lanl.gov/abs/1209.0716}{{\tt arXiv:1209.0716}}].

\bibitem{O'ShegOshinowo:2013naa}
{\bf NOvA} Collaboration, B.~O'Sheg~Oshinowo and H.~Freidsam, {\it {Survey of
  the NOvA Detectors at Fermilab}},  {\em \rm report
  FERMILAB-CONF-13-466-AD-PPD} (2013).

\bibitem{MINOS:1998nfa}
{\bf MINOS} Collaboration, {\it {Neutrino Oscillation Physics at Fermilab: The
  NuMI-MINOS Project}},  {\em \rm report NUMI-L-375} (1998).

\bibitem{Thron:1996vc}
{\bf MINOS} Collaboration, J.~L. Thron, {\it {The MINOS Long-Baseline Neutrino
  Oscillation Experiment}},  {\em \rm report NUMI-216} (1996) 1245--1250.

\bibitem{Papadimitriou:2014raa}
V.~Papadimitriou, R.~Andrews, J.~Hylen, T.~Kobilarcik, A.~Marchionni, {\em
  et.~al.}, {\it {Design of the LBNE Beamline}},  {\em \rm report
  FERMILAB-CONF-14-181-AD} (2014) TUOAA02.

\bibitem{Batell:2009di}
B.~Batell, M.~Pospelov, and A.~Ritz, {\it {Exploring Portals to a Hidden Sector
  Through Fixed Targets}},  {\em Phys.Rev.} {\bf D80} (2009) 095024,
  [\href{http://xxx.lanl.gov/abs/0906.5614}{{\tt arXiv:0906.5614}}].

\bibitem{deNiverville:2011it}
P.~deNiverville, M.~Pospelov, and A.~Ritz, {\it {Observing a light dark matter
  beam with neutrino experiments}},  {\em Phys.Rev.} {\bf D84} (2011) 075020,
  [\href{http://xxx.lanl.gov/abs/1107.4580}{{\tt arXiv:1107.4580}}].

\bibitem{deNiverville:2012ij}
P.~deNiverville, D.~McKeen, and A.~Ritz, {\it {Signatures of sub-GeV dark
  matter beams at neutrino experiments}},  {\em Phys.Rev.} {\bf D86} (2012)
  035022, [\href{http://xxx.lanl.gov/abs/1205.3499}{{\tt arXiv:1205.3499}}].

\bibitem{Batell:2014yra}
B.~Batell, P.~deNiverville, D.~McKeen, M.~Pospelov, and A.~Ritz, {\it
  {Leptophobic Dark Matter at Neutrino Factories}},
  \href{http://xxx.lanl.gov/abs/1405.7049}{{\tt arXiv:1405.7049}}.

\bibitem{Dharmapalan:2012xp}
{\bf MiniBooNE Collaboration} Collaboration, R.~Dharmapalan {\em et.~al.}, {\it
  {Low Mass WIMP Searches with a Neutrino Experiment: A Proposal for Further
  MiniBooNE Running}},  \href{http://xxx.lanl.gov/abs/1211.2258}{{\tt
  arXiv:1211.2258}}.

\bibitem{Goodman:2010yf}
J.~Goodman, M.~Ibe, A.~Rajaraman, W.~Shepherd, T.~M. Tait, {\em et.~al.}, {\it
  {Constraints on Light Majorana dark Matter from Colliders}},  {\em
  Phys.Lett.} {\bf B695} (2011) 185--188,
  [\href{http://xxx.lanl.gov/abs/1005.1286}{{\tt arXiv:1005.1286}}].

\bibitem{Bai:2010hh}
Y.~Bai, P.~J. Fox, and R.~Harnik, {\it {The Tevatron at the Frontier of Dark
  Matter Direct Detection}},  {\em JHEP} {\bf 1012} (2010) 048,
  [\href{http://xxx.lanl.gov/abs/1005.3797}{{\tt arXiv:1005.3797}}].

\bibitem{Goodman:2010ku}
J.~Goodman, M.~Ibe, A.~Rajaraman, W.~Shepherd, T.~M. Tait, {\em et.~al.}, {\it
  {Constraints on Dark Matter from Colliders}},  {\em Phys.Rev.} {\bf D82}
  (2010) 116010, [\href{http://xxx.lanl.gov/abs/1008.1783}{{\tt
  arXiv:1008.1783}}].

\bibitem{Fox:2011pm}
P.~J. Fox, R.~Harnik, J.~Kopp, and Y.~Tsai, {\it {Missing Energy Signatures of
  Dark Matter at the LHC}},  {\em Phys.Rev.} {\bf D85} (2012) 056011,
  [\href{http://xxx.lanl.gov/abs/1109.4398}{{\tt arXiv:1109.4398}}].

\bibitem{Goodman:2011jq}
J.~Goodman and W.~Shepherd, {\it {LHC Bounds on UV-Complete Models of Dark
  Matter}},  \href{http://xxx.lanl.gov/abs/1111.2359}{{\tt arXiv:1111.2359}}.

\bibitem{An:2012va}
H.~An, X.~Ji, and L.-T. Wang, {\it {Light Dark Matter and $Z'$ Dark Force at
  Colliders}},  {\em JHEP} {\bf 1207} (2012) 182,
  [\href{http://xxx.lanl.gov/abs/1202.2894}{{\tt arXiv:1202.2894}}].

\bibitem{Fox:2012ee}
P.~J. Fox, R.~Harnik, R.~Primulando, and C.-T. Yu, {\it {Taking a Razor to Dark
  Matter Parameter Space at the LHC}},  {\em Phys.Rev.} {\bf D86} (2012)
  015010, [\href{http://xxx.lanl.gov/abs/1203.1662}{{\tt arXiv:1203.1662}}].

\bibitem{An:2012ue}
H.~An, R.~Huo, and L.-T. Wang, {\it {Searching for Low Mass Dark Portal at the
  LHC}},  {\em Phys.Dark Univ.} {\bf 2} (2013) 50--57,
  [\href{http://xxx.lanl.gov/abs/1212.2221}{{\tt arXiv:1212.2221}}].

\bibitem{Fayet:2006sp}
P.~Fayet, {\it {Constraints on Light Dark Matter and U bosons, from psi,
  Upsilon, K+, pi0, eta and eta-prime decays}},  {\em Phys.Rev.} {\bf D74}
  (2006) 054034, [\href{http://xxx.lanl.gov/abs/hep-ph/0607318}{{\tt
  hep-ph/0607318}}].

\bibitem{Fayet:2007ua}
P.~Fayet, {\it {U-boson production in e+ e- annihilations, psi and Upsilon
  decays, and Light Dark Matter}},  {\em Phys.Rev.} {\bf D75} (2007) 115017,
  [\href{http://xxx.lanl.gov/abs/hep-ph/0702176}{{\tt hep-ph/0702176}}].

\bibitem{Fayet:2009tv}
P.~Fayet, {\it {Invisible Upsilon decays into Light Dark Matter}},  {\em
  Phys.Rev.} {\bf D81} (2010) 054025,
  [\href{http://xxx.lanl.gov/abs/0910.2587}{{\tt arXiv:0910.2587}}].

\bibitem{Graesser:2011vj}
M.~L. Graesser, I.~M. Shoemaker, and L.~Vecchi, {\it {A Dark Force for
  Baryons}},  \href{http://xxx.lanl.gov/abs/1107.2666}{{\tt arXiv:1107.2666}}.

\bibitem{Shoemaker:2011vi}
I.~M. Shoemaker and L.~Vecchi, {\it {Unitarity and Monojet Bounds on Models for
  DAMA, CoGeNT, and CRESST-II}},  {\em Phys.Rev.} {\bf D86} (2012) 015023,
  [\href{http://xxx.lanl.gov/abs/1112.5457}{{\tt arXiv:1112.5457}}].

\bibitem{Reece:2009un}
M.~Reece and L.-T. Wang, {\it {Searching for the light dark gauge boson in
  GeV-scale experiments}},  {\em JHEP} {\bf 0907} (2009) 051,
  [\href{http://xxx.lanl.gov/abs/0904.1743}{{\tt arXiv:0904.1743}}].

\bibitem{Bjorken:2009mm}
J.~D. Bjorken, R.~Essig, P.~Schuster, and N.~Toro, {\it {New Fixed-Target
  Experiments to Search for Dark Gauge Forces}},  {\em Phys.Rev.} {\bf D80}
  (2009) 075018, [\href{http://xxx.lanl.gov/abs/0906.0580}{{\tt
  arXiv:0906.0580}}].

\bibitem{Essig:2010xa}
R.~Essig, P.~Schuster, N.~Toro, and B.~Wojtsekhowski, {\it {An Electron Fixed
  Target Experiment to Search for a New Vector Boson A' Decaying to e+e-}},
  {\em JHEP} {\bf 1102} (2011) 009,
  [\href{http://xxx.lanl.gov/abs/1001.2557}{{\tt arXiv:1001.2557}}].

\bibitem{Essig:2010gu}
R.~Essig, R.~Harnik, J.~Kaplan, and N.~Toro, {\it {Discovering New Light States
  at Neutrino Experiments}},  {\em Phys.Rev.} {\bf D82} (2010) 113008,
  [\href{http://xxx.lanl.gov/abs/1008.0636}{{\tt arXiv:1008.0636}}].

\bibitem{Izaguirre:2013uxa}
E.~Izaguirre, G.~Krnjaic, P.~Schuster, and N.~Toro, {\it {New Electron
  Beam-Dump Experiments to Search for MeV to few-GeV Dark Matter}},  {\em
  Phys.Rev.} {\bf D88} (2013) 114015,
  [\href{http://xxx.lanl.gov/abs/1307.6554}{{\tt arXiv:1307.6554}}].

\bibitem{Morrissey:2014yma}
D.~E. Morrissey and A.~P. Spray, {\it {New Limits on Light Hidden Sectors from
  Fixed-Target Experiments}},  \href{http://xxx.lanl.gov/abs/1402.4817}{{\tt
  arXiv:1402.4817}}.

\bibitem{Izaguirre:2014dua}
E.~Izaguirre, G.~Krnjaic, P.~Schuster, and N.~Toro, {\it {Physics Motivation
  for a Pilot Dark Matter Search at Jefferson Laboratory}},  {\em Phys.Rev.}
  {\bf D90} (2014) 014052, [\href{http://xxx.lanl.gov/abs/1403.6826}{{\tt
  arXiv:1403.6826}}].

\bibitem{Nelson:1989fx}
A.~E. Nelson and N.~Tetradis, {\it {Constraints on a new vector boson coupled
  to baryons}},  {\em Phys.Lett.} {\bf B221} (1989) 80.

\bibitem{Carone:1994aa}
C.~D. Carone and H.~Murayama, {\it {Possible light U(1) gauge boson coupled to
  baryon number}},  {\em Phys.Rev.Lett.} {\bf 74} (1995) 3122--3125,
  [\href{http://xxx.lanl.gov/abs/hep-ph/9411256}{{\tt hep-ph/9411256}}].

\bibitem{Carone:1995pu}
C.~D. Carone and H.~Murayama, {\it {Realistic models with a light U(1) gauge
  boson coupled to baryon number}},  {\em Phys.Rev.} {\bf D52} (1995) 484--493,
  [\href{http://xxx.lanl.gov/abs/hep-ph/9501220}{{\tt hep-ph/9501220}}].

\bibitem{Dobrescu:2013cmh}
B.~A. Dobrescu and F.~Yu, {\it {Coupling-mass mapping of dijet peak searches}},
   {\em Phys.Rev.} {\bf D88} (2013), no.~3 035021,
  [\href{http://xxx.lanl.gov/abs/1306.2629}{{\tt arXiv:1306.2629}}].

\bibitem{Duerr:2013dza}
M.~Duerr, P.~Fileviez~Perez, and M.~B. Wise, {\it {Gauge Theory for Baryon and
  Lepton Numbers with Leptoquarks}},  {\em Phys.Rev.Lett.} {\bf 110} (2013)
  231801, [\href{http://xxx.lanl.gov/abs/1304.0576}{{\tt arXiv:1304.0576}}].

\bibitem{Aaltonen:2012jb}
{\bf CDF Collaboration} Collaboration, T.~Aaltonen {\em et.~al.}, {\it {A
  Search for dark matter in events with one jet and missing transverse energy
  in $p\bar{p}$ collisions at $\sqrt{s} = 1.96$ TeV}},  {\em Phys.Rev.Lett.}
  {\bf 108} (2012) 211804, [\href{http://xxx.lanl.gov/abs/1203.0742}{{\tt
  arXiv:1203.0742}}].

\bibitem{Aad:2011xw}
{\bf ATLAS Collaboration} Collaboration, G.~Aad {\em et.~al.}, {\it {Search for
  new phenomena with the monojet and missing transverse momentum signature
  using the ATLAS detector in $\sqrt{s}=7$ TeV proton-proton collisions}},
  {\em Phys.Lett.} {\bf B705} (2011) 294--312,
  [\href{http://xxx.lanl.gov/abs/1106.5327}{{\tt arXiv:1106.5327}}].

\bibitem{ATLAS:2012ky}
{\bf ATLAS Collaboration} Collaboration, G.~Aad {\em et.~al.}, {\it {Search for
  dark matter candidates and large extra dimensions in events with a jet and
  missing transverse momentum with the ATLAS detector}},  {\em JHEP} {\bf 1304}
  (2013) 075, [\href{http://xxx.lanl.gov/abs/1210.4491}{{\tt
  arXiv:1210.4491}}].

\bibitem{Chatrchyan:2012me}
{\bf CMS Collaboration} Collaboration, S.~Chatrchyan {\em et.~al.}, {\it
  {Search for dark matter and large extra dimensions in monojet events in $pp$
  collisions at $\sqrt{s}=7$ TeV}},  {\em JHEP} {\bf 1209} (2012) 094,
  [\href{http://xxx.lanl.gov/abs/1206.5663}{{\tt arXiv:1206.5663}}].

\bibitem{Khachatryan:2014rra}
{\bf CMS Collaboration} Collaboration, V.~Khachatryan {\em et.~al.}, {\it
  {Search for dark matter, extra dimensions, and unparticles in monojet events
  in proton-proton collisions at $\sqrt{s}$ = 8 TeV}},
  \href{http://xxx.lanl.gov/abs/1408.3583}{{\tt arXiv:1408.3583}}.

\bibitem{Aubert:2009ae}
{\bf BaBar Collaboration} Collaboration, B.~Aubert {\em et.~al.}, {\it {A
  Search for Invisible Decays of the Upsilon(1S)}},  {\em Phys.Rev.Lett.} {\bf
  103} (2009) 251801, [\href{http://xxx.lanl.gov/abs/0908.2840}{{\tt
  arXiv:0908.2840}}].

\bibitem{Ablikim:2007ek}
{\bf BES Collaboration} Collaboration, M.~Ablikim {\em et.~al.}, {\it {Search
  for the invisible decay of J / psi in psi(2S) --- pi+ pi- J / psi}},  {\em
  Phys.Rev.Lett.} {\bf 100} (2008) 192001,
  [\href{http://xxx.lanl.gov/abs/0710.0039}{{\tt arXiv:0710.0039}}].

\bibitem{Essig:2013vha}
R.~Essig, J.~Mardon, M.~Papucci, T.~Volansky, and Y.-M. Zhong, {\it
  {Constraining Light Dark Matter with Low-Energy $e^+e^-$ Colliders}},  {\em
  JHEP} {\bf 1311} (2013) 167, [\href{http://xxx.lanl.gov/abs/1309.5084}{{\tt
  arXiv:1309.5084}}].

\bibitem{Angloher:2014myn}
{\bf CRESST-II Collaboration} Collaboration, G.~Angloher {\em et.~al.}, {\it
  {Results on low mass WIMPs using an upgraded CRESST-II detector}},
  \href{http://xxx.lanl.gov/abs/1407.3146}{{\tt arXiv:1407.3146}}.

\bibitem{2012PhLB..711..264B}
J.~{Barreto}, H.~{Cease}, H.~T. {Diehl}, J.~{Estrada}, B.~{Flaugher},
  N.~{Harrison}, J.~{Jones}, B.~{Kilminster}, J.~{Molina}, J.~{Smith},
  T.~{Schwarz}, and A.~{Sonnenschein}, {\it {Direct search for low mass dark
  matter particles with CCDs}},  {\em Physics Letters B} {\bf 711} (May, 2012)
  264--269, [\href{http://xxx.lanl.gov/abs/1105.5191}{{\tt arXiv:1105.5191}}].

\bibitem{Agnese:2013jaa}
{\bf SuperCDMS Collaboration} Collaboration, R.~Agnese {\em et.~al.}, {\it
  {Search for Low-Mass Weakly Interacting Massive Particles Using
  Voltage-Assisted Calorimetric Ionization Detection in the SuperCDMS
  Experiment}},  {\em Phys.Rev.Lett.} {\bf 112} (2014), no.~4 041302,
  [\href{http://xxx.lanl.gov/abs/1309.3259}{{\tt arXiv:1309.3259}}].

\bibitem{Finkbeiner:2011dx}
D.~P. Finkbeiner, S.~Galli, T.~Lin, and T.~R. Slatyer, {\it {Searching for Dark
  Matter in the CMB: A Compact Parameterization of Energy Injection from New
  Physics}},  {\em Phys.Rev.} {\bf D85} (2012) 043522,
  [\href{http://xxx.lanl.gov/abs/1109.6322}{{\tt arXiv:1109.6322}}].

\bibitem{Galli:2011rz}
S.~Galli, F.~Iocco, G.~Bertone, and A.~Melchiorri, {\it {Updated CMB
  constraints on Dark Matter annihilation cross-sections}},  {\em Phys.Rev.}
  {\bf D84} (2011) 027302, [\href{http://xxx.lanl.gov/abs/1106.1528}{{\tt
  arXiv:1106.1528}}].

\bibitem{Hutsi:2011vx}
G.~Hutsi, J.~Chluba, A.~Hektor, and M.~Raidal, {\it {WMAP7 and future CMB
  constraints on annihilating dark matter: implications on GeV-scale WIMPs}},
  {\em Astron.Astrophys.} {\bf 535} (2011) A26,
  [\href{http://xxx.lanl.gov/abs/1103.2766}{{\tt arXiv:1103.2766}}].

\bibitem{Lin:2011gj}
T.~Lin, H.-B. Yu, and K.~M. Zurek, {\it {On Symmetric and Asymmetric Light Dark
  Matter}},  {\em Phys.Rev.} {\bf D85} (2012) 063503,
  [\href{http://xxx.lanl.gov/abs/1111.0293}{{\tt arXiv:1111.0293}}].

\bibitem{Zurek:2013wia}
K.~M. Zurek, {\it {Asymmetric Dark Matter: Theories, Signatures, and
  Constraints}},  {\em Phys.Rept.} {\bf 537} (2014) 91--121,
  [\href{http://xxx.lanl.gov/abs/1308.0338}{{\tt arXiv:1308.0338}}].

\bibitem{Boehm:2003hm}
C.~Boehm and P.~Fayet, {\it {Scalar dark matter candidates}},  {\em Nucl.Phys.}
  {\bf B683} (2004) 219--263,
  [\href{http://xxx.lanl.gov/abs/hep-ph/0305261}{{\tt hep-ph/0305261}}].

\bibitem{Steigman:2012nb}
G.~Steigman, B.~Dasgupta, and J.~F. Beacom, {\it {Precise Relic WIMP Abundance
  and its Impact on Searches for Dark Matter Annihilation}},  {\em Phys.Rev.}
  {\bf D86} (2012) 023506, [\href{http://xxx.lanl.gov/abs/1204.3622}{{\tt
  arXiv:1204.3622}}].

\bibitem{Alwall:2011uj}
J.~Alwall, M.~Herquet, F.~Maltoni, O.~Mattelaer, and T.~Stelzer, {\it {MadGraph
  5 : Going Beyond}},  {\em JHEP} {\bf 1106} (2011) 128,
  [\href{http://xxx.lanl.gov/abs/1106.0522}{{\tt arXiv:1106.0522}}].

\bibitem{Christensen:2008py}
N.~D. Christensen and C.~Duhr, {\it {FeynRules - Feynman rules made easy}},
  {\em Comput.Phys.Commun.} {\bf 180} (2009) 1614--1641,
  [\href{http://xxx.lanl.gov/abs/0806.4194}{{\tt arXiv:0806.4194}}].

\bibitem{Adamson:2011ku}
{\bf MINOS} Collaboration, P.~Adamson {\em et.~al.}, {\it {Active to sterile
  neutrino mixing limits from neutral-current interactions in MINOS}},  {\em
  Phys.Rev.Lett.} {\bf 107} (2011) 011802,
  [\href{http://xxx.lanl.gov/abs/1104.3922}{{\tt arXiv:1104.3922}}].

\bibitem{Soper:2014ska}
D.~E. Soper, M.~Spannowsky, C.~J. Wallace, and T.~M.~P. Tait, {\it {Scattering
  of Dark Particles with Light Mediators}},
  \href{http://xxx.lanl.gov/abs/1407.2623}{{\tt arXiv:1407.2623}}.

\bibitem{Adamson:2008qj}
{\bf MiniBooNE and Minos Collaboration} Collaboration, P.~Adamson {\em
  et.~al.}, {\it {First Measurement of $\nu_\mu$ and $\nu_e$ Events in an
  Off-Axis Horn-Focused Neutrino Beam}},  {\em Phys.Rev.Lett.} {\bf 102} (2009)
  211801, [\href{http://xxx.lanl.gov/abs/0809.2447}{{\tt arXiv:0809.2447}}].

\bibitem{Katori:2011uq}
{\bf MicroBooNE Collaboration} Collaboration, T.~Katori, {\it {MicroBooNE, A
  Liquid Argon Time Projection Chamber (LArTPC) Neutrino Experiment}},  {\em
  AIP Conf.Proc.} {\bf 1405} (2011) 250--255,
  [\href{http://xxx.lanl.gov/abs/1107.5112}{{\tt arXiv:1107.5112}}].

\bibitem{Admas:2013xka}
C.~Admas, C.~Andreopoulos, J.~Asaadi, B.~Baller, M.~Bishai, {\em et.~al.}, {\it
  {LAr1-ND: Testing Neutrino Anomalies with Multiple LAr TPC Detectors at
  Fermilab}},  {\em \rm report FERMILAB-PROPOSAL-1053} (2013).

\bibitem{Adams:2013qkq}
{\bf LBNE Collaboration} Collaboration, C.~Adams {\em et.~al.}, {\it {The
  Long-Baseline Neutrino Experiment: Exploring Fundamental Symmetries of the
  Universe}},  \href{http://xxx.lanl.gov/abs/1307.7335}{{\tt arXiv:1307.7335}}.

\end{thebibliography}\endgroup
\bibliographystyle{JHEP}


\end{document}